\documentclass[aoas]{imsart}
\usepackage{graphicx}
\usepackage{amssymb}
\usepackage{epstopdf}
\usepackage{natbib}
\usepackage{amsmath}
\usepackage{graphics}
\usepackage{bm}
\usepackage{color}
\usepackage{amsthm}
\usepackage{enumerate}
\usepackage{amsfonts}
\usepackage{algorithm}
\usepackage[noend]{algpseudocode}
\usepackage{bbm}

\newcommand{\ignore}[1]{}

\usepackage{float}
\floatstyle{ruled}
\newfloat{algo}{thp}{lop}
\floatname{algo}{Algorithm}
\newcommand{\brm}[1]{\bm{\mathrm{#1}}}
\newcommand{\E}{\mathbb{E}}
\newcommand{\Var}{\mathrm{Var}}

\newcommand{\tp}{\intercal}
\newcommand{\bms}{{\brm{s}}}
\newcommand{\bmu}{{\brm{u}}}
\newcommand{\bmv}{{\brm{v}}}
\newcommand{\bmp}{{\brm{p}}}
\newcommand{\bmm}{{\brm{m}}}
\newcommand{\bmD}{\brm{D}}
\newcommand{\bmgamma}{\bm{\gamma}}
\newcommand{\bmc}{\brm{c}}

\newcommand{\bme}{\brm{e}}
\newcommand{\bmS}{\brm{S}}
\newcommand{\bmV}{\brm{V}}
\newcommand{\bmtheta}{\bm{\theta}}

\newcommand{\bmH}{\brm{H}}

\newcommand{\cM}{\mathcal{M}}
\newcommand{\cS}{\mathcal{V}}
\newcommand{\acos}{\mathrm{arccos}}
\newcommand{\Exp}{\mathrm{Exp}}
\newcommand{\Log}{\mathrm{Log}}
\newcommand{\fs}{f_{S}}

\newcommand{\argmin}{\operatornamewithlimits{arg\ min}}

\newcommand{\suppref}[1]{{{#1}}}

\newtheorem{thm}{Theorem}

\newcommand{\linsp}{\renewcommand{\baselinestretch}{1}}
\newcommand{\linsps}{\renewcommand{\baselinestretch}{1}}

\newcommand{\revisedb}[1]{{#1}}
\newcommand{\revisedc}[1]{{#1}}
\newcommand{\revisedd}[1]{{#1}}
\newcommand{\reviseda}[1]{{#1}}
\newcommand{\jpeng}[1]{#1}
\newcommand{\rwong}[1]{#1}
\newcommand{\tlee}[1]{#1}
\newcommand{\dpaul}[1]{#1}

\begin{document}

\begin{frontmatter}
\title{Fiber Direction Estimation, Smoothing and Tracking in Diffusion MRI}
\runtitle{Fiber Direction Estimation in Diffusion MRI}

\begin{aug}
\author{\fnms{Raymond K. W.}
  \snm{Wong}\thanksref{a1}\ead[label=e1]{raywong@iastate.edu}},
\author{\fnms{Thomas C. M.}
  \snm{Lee}\thanksref{a2}\ead[label=e2]{tcmlee@ucdavis.edu}}
\author{\fnms{Debashis}
  \snm{Paul}\thanksref{a2}\ead[label=e3]{debpaul@ucdavis.edu}}
\author{\fnms{Jie}
  \snm{Peng}\thanksref{a2}\ead[label=e4]{jiepeng@ucdavis.edu}}
\and
\author{\fnms{}
  \snm{Alzheimer's Disease Neuroimaging Initiative}}

\runauthor{Wong et al.}

\affiliation{Iowa State University\thanksmark{a1} and
University of California at Davis\thanksmark{a2}}

\address{Department of Statistics \\
Iowa State University \\
2218 Snedecor Hall \\
Ames, IA 50011\\
\printead{e1}}

\address{Department of Statistics \\
University of California at Davis \\
4118 Mathematical Sciences Building \\
One Shields Avenue \\
Davis, CA 95616 \\
\printead{e2}\\
\phantom{E-mail:\ }\printead*{e3} \\
\phantom{E-mail:\ }\printead*{e4}}

\end{aug}

\maketitle

\begin{abstract}
Diffusion magnetic resonance imaging is an imaging technology designed to probe anatomical architectures of biological samples in an in vivo and non-invasive manner through measuring water diffusion.  \tlee{The contribution of this paper is threefold.  First it proposes a new method to identify and estimate multiple diffusion directions within a voxel through a new and identifiable parametrization of the widely used multi-tensor model.  Unlike many existing methods, this method focuses on the estimation of diffusion directions rather than the diffusion tensors.  Second, this paper proposes a novel direction smoothing method which greatly improves direction estimation in regions with crossing fibers.}  This smoothing method is shown to have excellent theoretical and empirical properties.  Lastly, this paper develops a fiber tracking algorithm that can handle multiple directions within a voxel.  The overall methodology is illustrated with simulated data and a data set collected for the study of Alzheimer's disease by the Alzheimer's Disease Neuroimaging Initiative (ADNI).

\end{abstract}


\end{frontmatter}

\section{Introduction}
Diffusion magnetic resonance imaging (dMRI) is an in vivo and non-invasive
medical imaging technology that uses water diffusion as a proxy to probe
anatomical structures of biological samples.   The most important application
of dMRI is to  reconstruct white matter fiber tracts \revisedc{in brain} -- large \revisedd{axonal}
bundles with similar destinations.
In white matter,
  water diffusion appears to be anisotropic as water tends to diffuse faster along the fiber bundles. Therefore,  white matter fiber structures can be deduced from the diffusion characteristics of water.
   Mapping white matter fiber tracts is of great importance in the study of structural organization of neuronal networks and the understanding of brain functionality
\citep{Mori07, Sporns11}. Moreover, dMRI also \revisedd{has} many
  clinical applications, including detecting brain abnormality in white matter
  due to axonal loss or deformation, which are thought to be \revisedc{related to}
  many neuron degenerative diseases including Alzheimer's disease, \revisedd{and also in} surgical
  planning by resolving complex neuronal connections  between white and gray
  matter \citep{nimsky2006implementation}.

  \jpeng{dMRI techniques sensitize  signal intensity with the amount of water
    diffusion by applying pulsed magnetic gradient fields on the sample.
    Specifically, water diffusion along the gradient field direction leads to signal
    loss and the amount of loss at a voxel equals to the summation (across
    locations within the voxel) of the sinusoid waves with shifted signal phases
    weighted by the proton density at their respective locations. In other words,
    signal loss (referred to as diffusion weighted signal) is the inverse Fourier
    transform of the diffusion probability density function of water molecules and thus can be used to
    recover water diffusion characteristics. The amount of signal loss is also
    influenced by various experimental parameters including  the gradient field
    intensity (the stronger, the more loss), the  duration of gradient fields (the
    longer, the more loss), etc. Their effects are aggregatively reflected by an
    experimental  parameter  called the ``$b$-value" which is often fixed throughout
    the experiment (though multiple $b$-values are used in Q-space imaging). Since
    only water motion along the gradient field direction can be detected, multiple
    gradient directions need to be applied \citep{Mori07}.}

\begin{figure}[htpb]
\begin{center}
  \includegraphics[height=5cm]{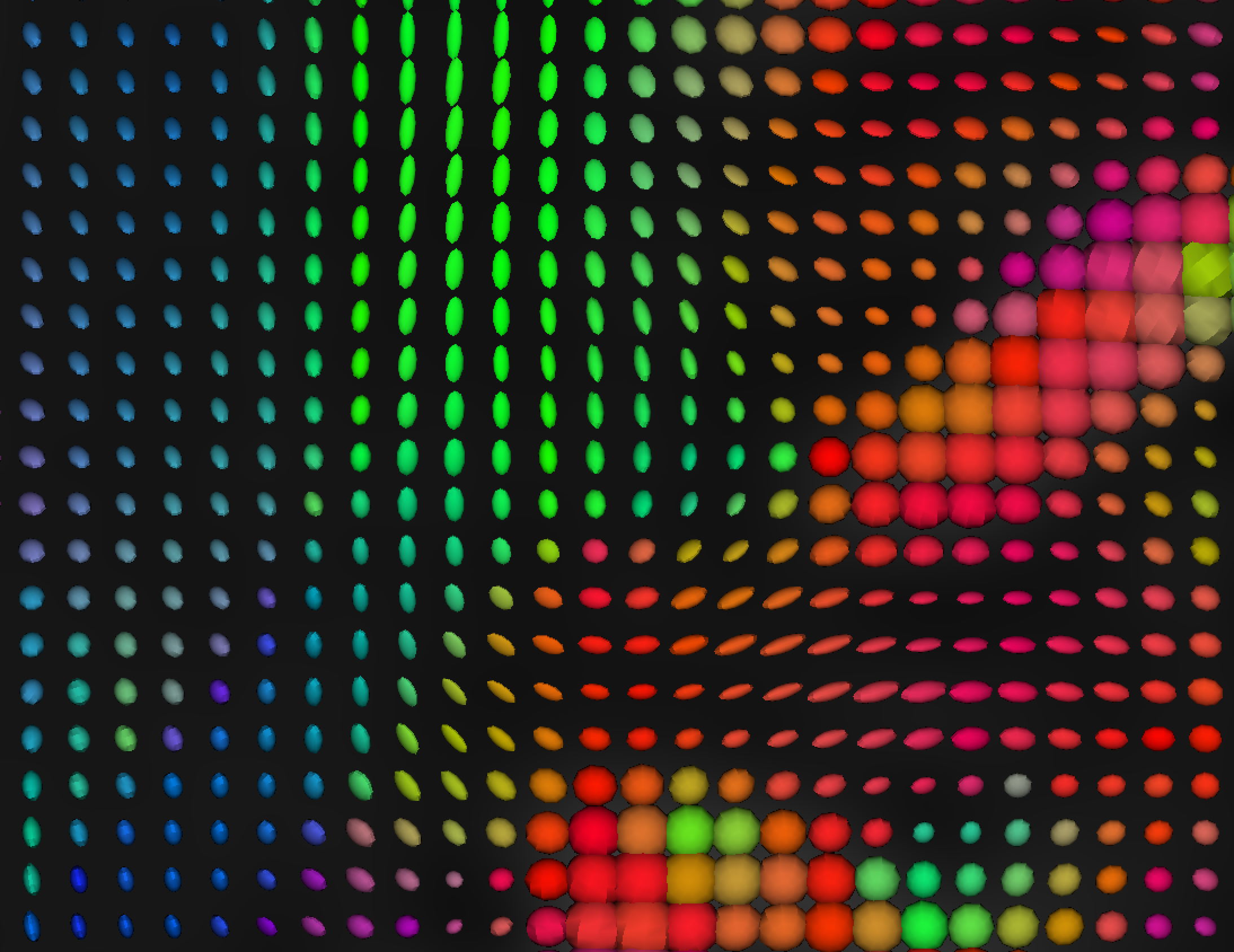}
\end{center}
  \caption{\rwong{An example of a tensor map on a 2D grid, where each diffusion tensor is represented by an elliposid.}}\label{fig:tensormap}
\end{figure}

    \jpeng{In its raw form, dMRI
    provides diffusion weighted signal measurements on a 3D spatial grid (of the
    sample) along a set of predetermined gradient directions}
\citep{Bammer-Holdsworth-Veldhuis09, Beaulieu02, Chanraud-Zahr-Sullivan10,
  Mukherjee-Berman-Chung08}. For example, a typical data set from the Alzheimer's Disease Neuroimaging Initiative (ADNI) has diffusion measurements along $41$ gradient directions for each voxel on a $256 \times 256 \times 59$ 3D grid of the brain. The first step of dMRI analysis
  is to summarize these measurements into estimates of water diffusion at each voxel. A popular model for water diffusion is the so called single tensor model where the diffusion process is \revisedd{modeled} as a 3D Gaussian process described by a $3 \times 3$ positive definite matrix, referred to as a diffusion tensor; see \citet{Mori07} for an introduction to diffusion tensor imaging (DTI) techniques.
  \rwong{Figure~\ref{fig:tensormap} depicts a tensor map on a 2D grid, where each diffusion tensor is represented by an ellipsoid, estimated from diffusion weighted measurements from an ADNI data set using a single tensor model.}
  One then extracts the
  \revisedd{local diffusion direction as the principal eigenvector of the (estimated) diffusion tensor}
 at each voxel and reconstructs the white matter fiber tracts by computer aided tracking algorithms via a process called tractography \citep{Basser-Pajevic-Pierpaoli00}.

However, DTI cannot resolve multiple fiber populations with distinct orientations, i.e., crossing fibers, within a voxel since a tensor only has one principal direction. Consequently, in crossing fiber regions, estimated diffusion tensors may lead to low anisotropy estimation or oblate tensor estimation.  Poor tensor estimation results in poor direction estimation which adversely affects fiber reconstruction; e.g., early termination of or biased fiber tracking.

In order to resolve intravoxel orientational heterogeneity, several approaches have been proposed.  \citet{Tuch-Reese-Wiegell02} propose a multi-tensor model which assumes a finite number of homogeneous fiber directions within a voxel.  However, it has been shown that the parameters in the multi-tensor model are not identifiable \citep{Scherrer-Warfield10}.
\jpeng{Imaging techniques such as Q-ball and Q-space and the corresponding nonparametric methods have also been proposed}
\citep{Tuch04, Descoteaux-Angelino-Fitzgibbons07}.  However such methods often require high angular resolution  diffusion imaging (HARDI) \citep{Tuch-Reese-Wiegell02, Hosey-Williams-Ansorge05} where a large number of gradients is sampled.
\revisedc{In light of these facts},  the goal of this paper is to develop a new fiber \tlee{direction estimation}
and tracking method that can handle crossing fibers without requiring any high
resolution techniques.  The proposed method, named DiST, short for
\revisedd{\textbf{Di}ffusion Direction \textbf{S}moothing and \textbf{T}racking},
is completely \revisedd{automated} and improves existing methods in several aspects.
\revisedc{Particularly, it} is applicable either when there is a large number
of gradient directions (\revisedc{as} in the HARDI setting) or when only a
relatively small number of gradient directions \revisedc{are} available (\revisedc{as} in most clinical
settings).

The DiST method can be divided into three major steps.

{\bf Step 1:}
\revisedd{Estimate} the tensor directions within each voxel under a multi-tensor model.  A new parametrization is proposed which makes the tensor directions identifiable.  An efficient and numerically stable computational procedure is developed to obtain the maximum likelihood (ML) estimate of the tensor directions.  \tlee{Here we highlight that, this method focuses on the estimation of the tensor directions rather than the actual tensors themselves.}

{\bf Step 2:}
\revisedd{Using the voxel-wise tensor direction estimates from Step~1 as input,}
a new direction smoothing procedure is applied to further improve the diffusion direction estimates by borrowing information from neighboring voxels.  A distinctive and unique feature of this smoothing procedure is that it handles crossing fibers through the clustering of directions into homogeneous groups.  We note that, although various tensor smoothing methods have been proposed
\citep[e.g.,][]{Pennec-Fillard-Ayache06,Arsigny-Fillard-Pennec06,Fillard-Pennec-Arsigny07,Fletcher-Joshi07,Yuan-Zhu-Lin12,Carmichael-Chen-Paul13},
little work has been done on direct diffusion direction smoothing.
\revisedb{
One notable exception is \citet{Schwartzman-Dougherty-Taylor08}, which
harnesses diffusion
directions directly to construct a map of test statistics for detecting differences
between diffusion direction maps from two groups of subjects,
\revisedc{while the spatial smoothness of the test statistics is being considered.}
Also note that approaches to averaging unsigned directions in the real
projective space are known in the directional statistics literature.
}

{\bf Step 3:}
Lastly, a fiber tracking algorithm is applied to reconstruct fiber tracts using the smoothed diffusion direction estimates obtained in Step~2.  This tracking algorithm is designed to explicitly allow for multiple directions within a voxel.

We apply DiST to an ADNI data set measured on a healthy elderly person with 41-direction dMRI scan on a 3 Tesla GE
Medical Systems MRI scanner. ADNI is a longitudinal study (since 2005) that collects serial
MRI, cognitive assessments, and numerous additional measurements approximately twice per year
from hundreds of elderly individuals spanning a range from cognitive health to clinically-diagnosed Alzheimer's
disease. We also examine DiST using simulated data sets which mimic the most commonly encountered experimental situations in terms of number of gradient directions and signal to noise ratio. \tlee{DiST is shown} to lead to  superior
results than those based on the single tensor model in the simulation study, as well as more biologically sensible results in the real data application.

The rest of the paper is organized as follows.  Section~\ref{sec:models} provides background material for some common tensor models.  The proposed methods for tensor direction estimation, smoothing of estimated directions, and fiber tracking are presented in, respectively, Sections~\ref{sec:voxelwise}, \ref{sec:smoothing} and~\ref{sec:track}.
Section~\ref{sec:sim1} summarizes simulation results.  
The application to an ADNI data set is presented in Section~\ref{sec:real}.  Section~\ref{sec:discuss} provides some concluding remarks, while additional \rwong{simulation} results and technical details are collected in an online Supplemental Material.

\section{Tensor models}\label{sec:models}
Suppose dMRI measurements are made on $N$ voxels on a 3D grid representing a brain.  For each voxel, we have measurements of diffusion weighted signals along a fixed set (i.e., the same for all voxels) of unit-norm gradient vectors $\mathcal{U}=\{\bmu_i:i=1,\dots,m\}$.
We write the set of measurements as $\{S(\bms, \bmu): \bmu\in \mathcal{U} \}$, where $\bms$ is the 3D coordinate of the center of this voxel.

Assuming Gaussian diffusion, the noiseless signal intensity is given by \citep[e.g.,][]{Mori07}
\[
\bar{S} (\bms, \bmu) = S_0(\bms) \exp\left\{-b\bmu^\tp {\bmD}(\bms) \bmu\right\},
\]
where $S_0(\bms)$ is the non-diffusion-weighted intensity, $b>0$ is an
experimental constant referred to as the $b$-value and $\bmD(\bms)$ is a
$3\times 3$ covariance matrix referred to as the diffusion tensor.  This model
is called the single tensor model and suits for the case of at most one
dominant diffusion direction within a voxel.

Although the single tensor model is the most widely used tensor model in practice, it is not suitable for crossing fiber regions.  To deal with crossing fibers, this model has been extended to a multi-tensor model
\citep[e.g.,][]{Tuch02,
  Behrens-Woolrich-Jenkinson03,
  Behrens-Berg-Jbabdi07,
  Tabelow-Voss-Polzehl12}:
\begin{align}
\bar{S}(\bms, \bmu) = S_0(\bms)\sum^{J(s)}_{j=1} p_j(\bms) \exp\left\{-b\bmu^\tp
  {\bmD}_j(\bms) \bmu\right\}, \label{eqn:truedti}
\end{align}
where $\sum^{J(\bms)}_{j=1} p_j(\bms)=1$ and $p_j(\bms)>0$ for $j=1,\dots,J(\bms)$.  Here $J(\bms)$ represents the number of fiber populations and $p_j(\bms)$'s denote weights of the corresponding fibers.

\section{Voxel-wise estimation of diffusion directions}
\label{sec:voxelwise}
One important goal of \revisedc{dMRI} studies is to estimate principal diffusion directions, referred to as diffusion directions hereafter, at each voxel.  They may be interpreted as tangent directions along fiber bundles at the corresponding voxel.  The estimated diffusion directions are then used as an input for tractography algorithms to reconstruct fiber tracts.  This section explores the diffusion direction estimation within a single voxel.  For notational simplicity, dependence on voxel index $\bms$ is temporarily dropped.  Moreover, for ease of exposition, we assume that $\sigma$ and $S_0(\bms)$ are known {and delay the discussion of their estimation to Section~\ref{sec:real}.

Under the single tensor model, various methods for tensor estimation have been proposed including linear regression, nonlinear regression and ML estimation; e.g., see \citet{Carmichael-Chen-Paul13} for a comprehensive review.  Then diffusion directions are derived as principal eigenvectors of (estimated) diffusion tensors.  However, for the estimation of multi-tensor models, severe computational issues have been observed and additional prior information and additional assumptions are usually imposed to tackle these issues.  For instance, \citet{Behrens-Woolrich-Jenkinson03,Behrens-Berg-Jbabdi07} use shrinkage priors and \citet{Tabelow-Voss-Polzehl12} assume all tensors to be axially symmetric (i.e., the two minor eigenvalues are the same) and have the same set of eigenvalues.  \citet{Scherrer-Warfield10} show that the multi-tensor model is indeed non-identifiable \tlee{in the sense that there exist multiple parameterizations that are observationally equivalent}.  These authors suggest to use multiple $b$-values in data acquisition to make the model identifiable.  However, due to practical limitations, most of the current dMRI studies are obtained under a fixed $b$-value and so render their suggestion inapplicable.  Below we show that the identifiability issue does not prevent one from estimating the diffusion directions and so neither strong assumptions nor special experimental settings are necessary if one is only interested in diffusion directions rather than the diffusion tensors themselves.

\subsection{Identifiability of multi-tensor model}
Model~(\ref{eqn:truedti}) can be re-written as
\[
  \bar{S}(\bmu) = S_0 \sum^{J}_{j=1} p_j a_j \exp \left\{
  -b \bmu^\tp\left( {\bmD}_j + \frac{\log a_j}{b} \brm{I}_3\right) \bmu
  \right\},
\]
where $a_j>0$ for $j=1,\dots,J$ such that $p_ja_j>0$, $\bmD_j +
(\log a_j/b)\brm{I}_3$ is positive definite and $\sum^J_{j=1}{p_ja_j}=1$. When
$J=2$, one can easily derive the explicit conditions for $a_j$ to fulfill
these criteria, and see that there are infinite sets of such $a_j$'s.
However, note that $\bmD_j +
(\log a_j/b)\brm{I}_3$ shares the same set of eigenvectors with ${\bmD}_j$. Thus,
one may still be able to estimate diffusion directions, which correspond to the
major eigenvectors of the tensors. This motivates us to consider estimating
diffusion directions directly instead of the tensors themselves.

Now we assume that ${\bmD}_j$'s are axially symmetric; that is, the two minor eigenvalues of ${\bmD}_j$ are equal.
This is a common assumption
for modeling dMRI data and it implies that
diffusion is symmetric around the principal
diffusion direction
\citep{Tournier-Calamante-Gadian04, Tournier-Calamante-Connelly07}.
By not differentiating the two minor eigenvectors, we obtain a
clear meaning of diffusion direction.  In addition, this reduces the number of
unknown parameters by one \reviseda{for each tensor in
the multiple tensor model} and thus facilitates estimation.
In the following, we propose a new parametrization of the multi-tensor model which is identifiable and thus can be used for direction estimation.

Write $\cM$ as the space of the unit principal eigenvector; i.e., the  3D unit sphere with equivalence relation
$\bmm \sim -\bmm$.
Let $\alpha_j\ge0$, $\xi_j>0$ and ${\bmm}_j\in\cM$
be the difference between the larger and smaller eigenvalue, smaller eigenvalue
and the standardized principal eigenvector of ${\bmD}_j$, respectively.
Since $\bmD_j =\alpha_j {\bmm}_j {\bmm}_j^\tp + \xi_j \brm{I}_3 $,
model~(\ref{eqn:truedti}) becomes
\begin{align}
  \bar{S}(\bmu) &= S_0 \sum^J_{j=1}p_j \exp \left\{ -b \bmu^\tp \left(
    \alpha_j {\bmm}_j {\bmm}_j^\tp + \xi_j \brm{I}_3\right) \bmu \right\}\nonumber\\
&= S_0 \sum^J_{j=1} \tau_j \exp \left\{ -b \alpha_j (\bmu^\tp
  {\bmm}_j)^2\right\},\label{eqn:newpar}
\end{align}
where $\tau_j = p_j \exp(-b\xi_j) \in (0,1)$.
 From the above, one can see that
$p_j$ and $\xi_j$ are not simultaneously identifiable, so
we cannot estimate the tensors.
However, as stated in the following theorem, $\tau_j, \alpha_j, \bmm_j^\tp$ are identifiable and hence we can estimate the principal diffusion directions $\bmm_j$'s.

\begin{thm}\label{thm:identi}
Under model~(\ref{eqn:newpar}), \tlee{for any arbitrary $J$}, the parameters $\bmgamma=(\bmgamma_1^\tp,\dots,\bmgamma_J^\tp)^\tp$ are identifiable,
where
$\bmgamma_j=(\tau_j, \alpha_j, \bmm_j^\tp)^\tp$ for $j=1, \dots, J$.
\end{thm}
\revisedb{The proof of this theorem can be found in \suppref{Section~S5.1} of the Supplemental Material. Note that,
  \revisedc{compared to the model in} \citet{Tabelow-Voss-Polzehl12},
  \revisedc{model (\ref{eqn:newpar})} allows for different eigenvalues and shapes of the
  tensors within a voxel, \revisedc{and thus is much more flexible}.}

\subsection{Parameter estimation using \revisedc{maximum likelihood (ML)}}
We first consider the case when $J$ is known and delay the selection of $J$ to Section~\ref{sec:J}.
\revisedb{By assuming Gaussian additive noise on both real and imaginary parts of the
  \revisedc{complex} signal, the observed signal intensity can be modeled as \reviseda{\citep[see, e.g.,][]{Zhu-Zhang-Ibrahim07}}
\[
S(\bmu) = \| \bar{S}( \bmu) \bm{\phi}( \bmu) + \sigma
\bm{\epsilon}(\bmu)\|,
\]
where $\bar{S}( \bmu)$ is the intensity of the noiseless signal, $\bm{\phi}(\bmu)$ is a unit vector in $\mathbb{R}^2$ representing the phase of the signal, $\bm{\epsilon}(\bmu)$ is the noise random variable following $\mathcal{N}_2(\brm{0},\brm{I}_2)$ and $\sigma>0$ denotes the noise level.
Note that both $\phi$ and $\bm{\epsilon}$ \revisedc{may} depend on $\bms$.
The observed signal intensity then follows a Rician distribution \citep{Gudbjartsson-Patz95}:
\[
S(\bmu) \sim \mathrm{Rician}(\bar{S}(\bmu),\sigma).
\]
Moreover, we assume the noise $\bm{\epsilon}(\bmu)$'s are independent across different voxels and gradient directions.}

Under the Rician noise assumption, the log-likelihood of $\bmgamma$ in model~(\ref{eqn:newpar}) is:
\begin{align}
l(\bmgamma) &= \sum_{\bmu\in\mathcal{U}}\log\left[
  \frac{S(\bmu)}{\sigma^2} \exp \left\{ -\frac{S^2(\bmu) +
      \bar{S}^2(\bmu)}{2\sigma^2}  \right\} I_0 \left\{\frac{S(\bmu)
      \bar{S}(\bmu)}{\sigma^2}\right\} \right] \notag\\
&= \sum_{\bmu\in\mathcal{U}} \left[ \log\left\{\frac{S(\bmu)}{\sigma^2}\right\}
  -\frac{S^2(\bmu) + \bar{S}^2(\bmu)}{2\sigma^2} + \log I_0 \left\{\frac{S(\bmu)
      \bar{S}(\bmu)}{\sigma^2}\right\} \right], \label{eqn:likelihood}
\end{align}
where $I_0(x) = \int^{\pi}_0 \exp(x\cos\phi) d\phi/ \pi$ is the zeroth order modified Bessel function of the first kind.
The ML estimate is obtained through maximizing~(\ref{eqn:likelihood}). Although the above new parametrization avoids the identifiability issue, the likelihood function usually has multiple local maxima, which makes the computation of ML estimate difficult and unstable.

The method that we used to overcome this issue can be briefly described as follows.  We first develop an approximation of model~(\ref{eqn:newpar}) whose likelihood can be globally maximized via a grid search.  We utilize the geometry of the problem so that the grid search can be done \revisedc{efficiently}.  Then we use the ML estimate of this approximated model as the initial value in a gradient method to obtain the ML estimate of model~(\ref{eqn:newpar}).  This method provides very reliable estimates.  To speed up the pace of this article, its full description is given in \suppref{Section~S1} of the Supplemental Material.

\subsection{Selecting the number of tensor components $J$}\label{sec:J}
Common model selection methods can be applied to select the number of components $J$.  Results from extensive numerical experiments suggest that the Bayesian information criterion (BIC) \citep{Schwarz78} is good choice; see \suppref{Section~S2} of the Supplemental Material.

\reviseda{
  Under model~(\ref{eqn:newpar}),
  each \revisedc{tensor} corresponds to four free scalar parameters since
  $\brm{m}_j$ is characterized by two free scalar parameters.
}
The BIC for a model with $I$ \revisedc{tensors} is
\begin{equation}
  \mathsf{BIC}(I) = -2l(\hat{\bmgamma}(I)) + 4I\log(m), \label{eqn:bic}
\end{equation}
where $m$ is the number of gradient directions and $\hat{\bmgamma}(I)$ is the ML estimate of $\bmgamma$ under $I$ tensors.
Then $J$ is chosen as
$
  \hat{J} = \mathrm{argmin}_{I \in \{1, \dots, \tilde{I} \}}
  \mathsf{BIC}(I),
$
where $\tilde{I}$ is a pre-specified upper bound for the number of components.
Based on our experience, $\tilde{I}=4$ is a reasonable choice.

In practice, there are voxels with no major diffusion directions.
\revisedc{This corresponds to the case where there is only one isotropic tensor.}
In the case of isotropic tensor,
(\ref{eqn:newpar}) reduces to
$
  \bar{S}(\bmu) = \bmS_0\tau_1.
$
Thus there is only one parameter $\tau_1$. We write
the corresponding likelihood function as $\tilde{l}$
and denote the ML estimate of $\tau_1$ by $\hat{\tau}_1$,
which can be obtained by a generic gradient method.
The corresponding
BIC criterion is
\[
  \mathsf{BIC}(0) = -2 \tilde{l}(\hat{\tau}_1) + \log(m),
\]
where 0 represents no diffusion direction.
Combined with the previous BIC formulation~(\ref{eqn:bic}), one has a comprehensive
model selection rule, which handles voxels with from zero to up to $\tilde{I}$ (here 4) fiber
populations.

\revisedb{In practice, we follow the convention and use
fractional anisotropy (FA) \citep[see, e.g.,][]{Mori07},
\begin{equation}
   FA = \sqrt{\frac{ (\lambda_1-\lambda_2)^2 + (\lambda_2-\lambda_3)^2 +
   (\lambda_3-\lambda_1)^2 }{{2(\lambda_1^2 + \lambda_2^2 + \lambda_3^2})}}, \label{eqn:FA}
\end{equation}
where $\lambda_1$, $\lambda_2$ and $\lambda_3$ are the eigenvalues of the corresponding tensor in the single tensor model,
\revisedc{to conduct an initial screening} to speed up the whole procedure.
The FA value lies between zero and one and the larger it is, the more anisotropic the water diffusion is at the corresponding voxel.
Thus we
first remove voxels with very small FA values and then apply the BIC
approach over those suspected anisotropic
voxels.}
\rwong{Note that such removal is mainly for reducing computational cost as a typical
dMRI data set consists of hundreds of thousands of voxels. From our experiences,
this has little effect on the final tracking results.
We also note that the proposed framework including selection of $J$ can be applied
without such removal if enough computational resources are available.}

We summarize our voxel-wise estimation procedure in \suppref{Algorithm~S2} in the Supplemental Material.  A simulation study is conducted and the corresponding results are presented in \suppref{Section~S2} of the Supplemental Material.  These numerical results suggest that our voxel-wise estimation procedure provides extremely stable and reliable results under various settings.


\section{Spatial smoothing of diffusion directions}
\label{sec:smoothing}
Although model~(\ref{eqn:newpar}) provides a better modeling than the single tensor model for crossing fiber regions, it also leads to an increase in the number of parameters and thus the variability of the estimates.
To further improve estimation, we consider borrowing information from neighboring voxels and develop a novel smoothing technique for diffusion directions.

\revisedc{In many brain regions, it is reasonable to model the fiber tracts as smooth curves at the resolution of voxels in dMRI ($\sim$ 2mm).
Therefore, we shall assume that the tangent directions of fiber bundles change smoothly.  This leads to the spatial smoothness of diffusion directions that belong to the same fiber bundle.}

\subsection{Smoothing along a single fiber}\label{sec:single_smooth}

This subsection considers the simpler situation where there is only one homogeneous population of diffusion directions; i.e., there is only one single fiber bundle without crossing.
Write $T$ as the total number of estimated diffusion directions from all voxels and $\{\hat{\bmm}_k: k=1,\dots, T\}$ as the set of all estimated diffusion directions.  Also write $\bms_k$ as the corresponding voxel location associated with $\hat{\bmm}_k$.  Note that some $\bms_k$'s share the same value, as some voxels contain multiple estimated directions.
Following the idea of kernel smoothing on Euclidean space \citep[e.g.,][]{Fan-Gijbels96}, the smoothing estimate at voxel $\bms_0$ is defined as a weighted Karcher mean of the neighboring direction vectors:
\begin{equation}
  \argmin_{\bmv\in\cM} \sum^T_{i=1}
   w_id^{*2}(\hat{\bmm}_i,{\bmv}), \label{eqn:general_smooth}
\end{equation}
where $w_i = K_{\bmH}(\bms_i-\bms_0)$'s are spatial weights
and the metric $d^*$ is defined as
\begin{equation}
  d^{*} (\bmu,\bmv) = \mathrm{arccos}(|\bmu^\tp \bmv|), \quad
\bmu, \bmv \in \cM; \label{eqn:dist2}
\end{equation}
i.e., $d^{*} (\bmu,\bmv)$ is the acute angle between $\bmu$ and $\bmv$.
The weights $w_i$'s place more emphasis on spatially closer observations.
Here $K_{\bmH}(\cdot)=|\bmH|^{-1/2}K({\bmH}^{-1/2}\cdot)$ with $K(\cdot)$ as a
3D kernel function satisfying $\int K(\bms) d\bms = 1$, and
$\bmH$ is a $3 \times 3$ bandwidth matrix.  In our numerical work, we choose
$K(\cdot)$ as the standard Gaussian density, and set $\bmH = h\brm{I}_3$, where
$h$ is \revisedd{chosen using} the cross-validation (CV) approach described in
\suppref{Section~S3} of the Supplemental Material.
\revisedb{We adopt the leave-one-out CV idea to
develop an ordinary CV score and two robust CV scores.
\rwong{Their practical performances are reported
in \suppref{Section~S6} of the Supplemental Material.
}}

\subsection{Smoothing over multiple fibers}\label{sec:mult_smooth}
When there are crossing fibers in a voxel $\bms_0$, the above smoothing procedure will not work well.  To \revisedc{address} this issue, we first cluster the neighboring estimated directions of $\bms_0$ into \revisedc{groups} that correspond to different fiber populations.  Then we apply the above smoothing procedure to each individual \revisedc{cluster}.  This subsection describes this procedure in details.

First we define neighboring voxels for $\bms_0$.  We begin with
computing the spatial weights defined in Section~\ref{sec:single_smooth}.  We
then remove those voxels with weights smaller than a threshold.  By filtering
out these voxels, we obtain tighter and better separated clusters of directions.
Moreover,
such voxels have little effects on smoothing due to their small weights.  The
artificial data set displayed in Figure~\ref{fig:separate} provides an
illustrative example.  Each black dot in the left panel represents an
estimated direction (from the center of the sphere).  In the middle panel, the
size of each dot is proportional to its spatial weight in equation
(\ref{eqn:general_smooth}).  Lastly, the right
panel shows all dots with spatial weights larger than a threshold.  Notice that
such a trimming operation leads to two obvious clusters of directions, which makes
the subsequent task of clustering the directions much easier.

\begin{figure}[htpb]
\[
  \includegraphics[height=3cm]{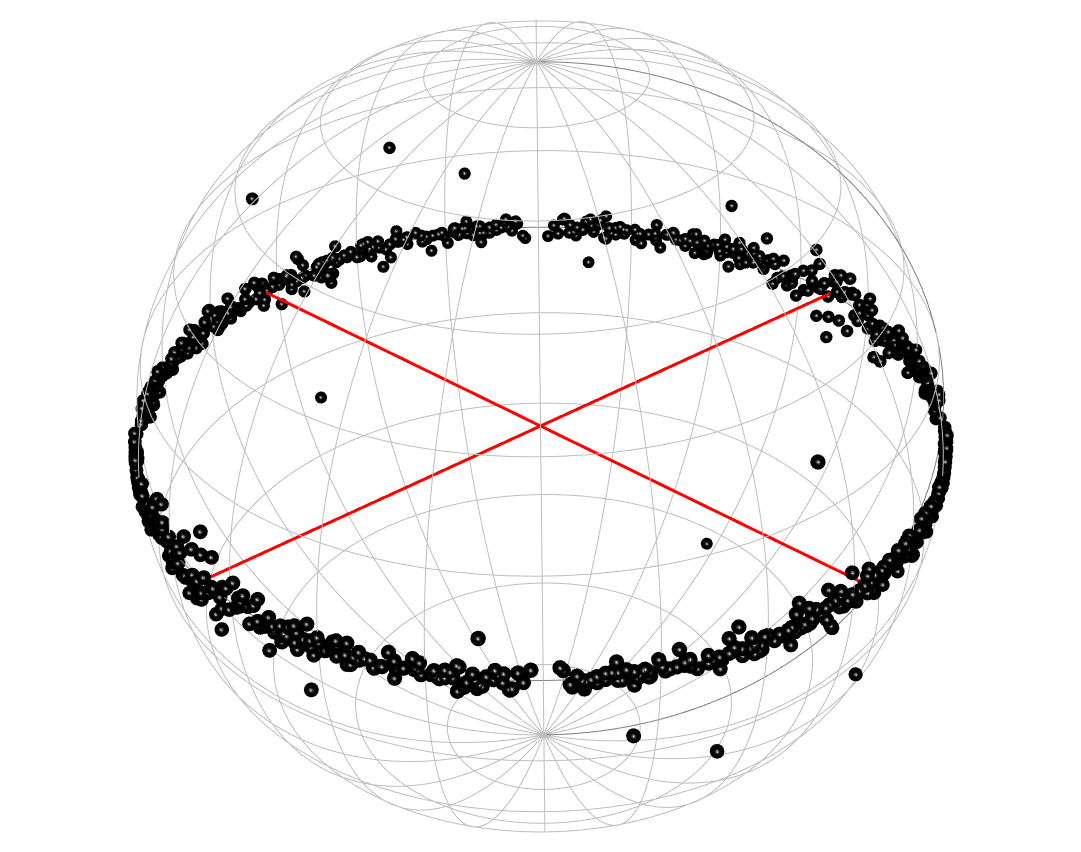}
  \includegraphics[height=3cm]{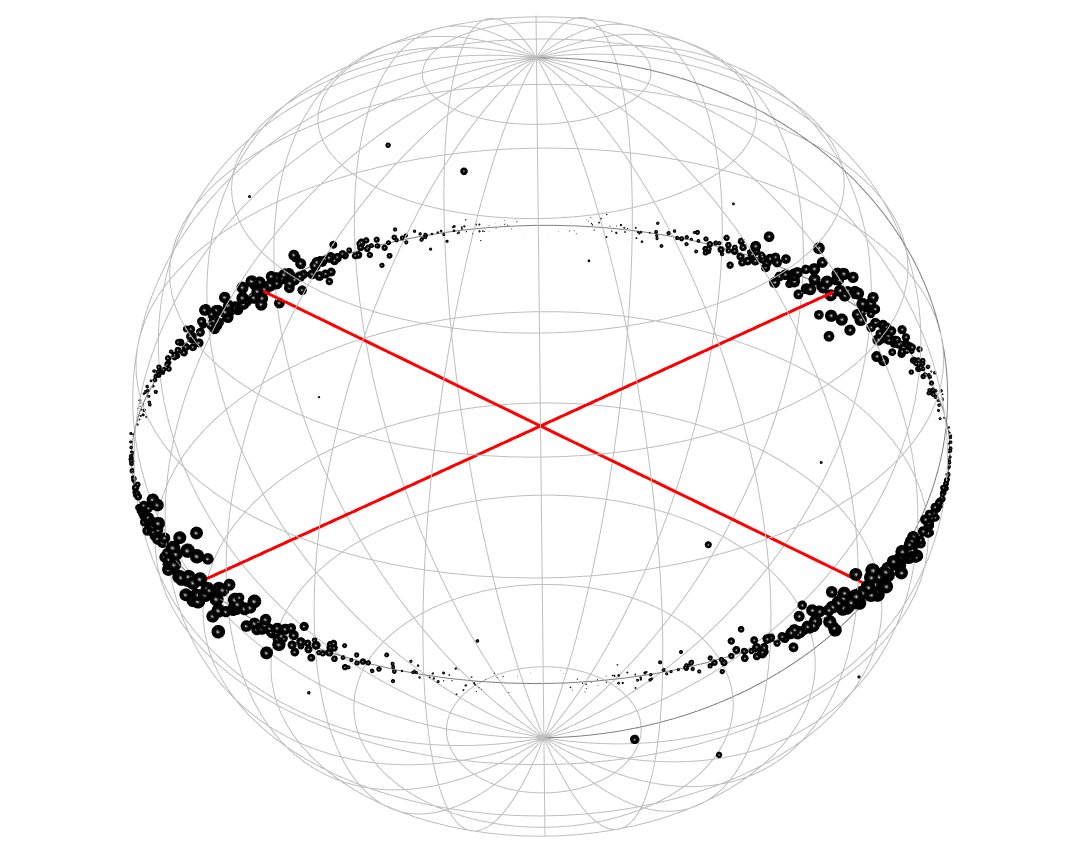}
  \includegraphics[height=3cm]{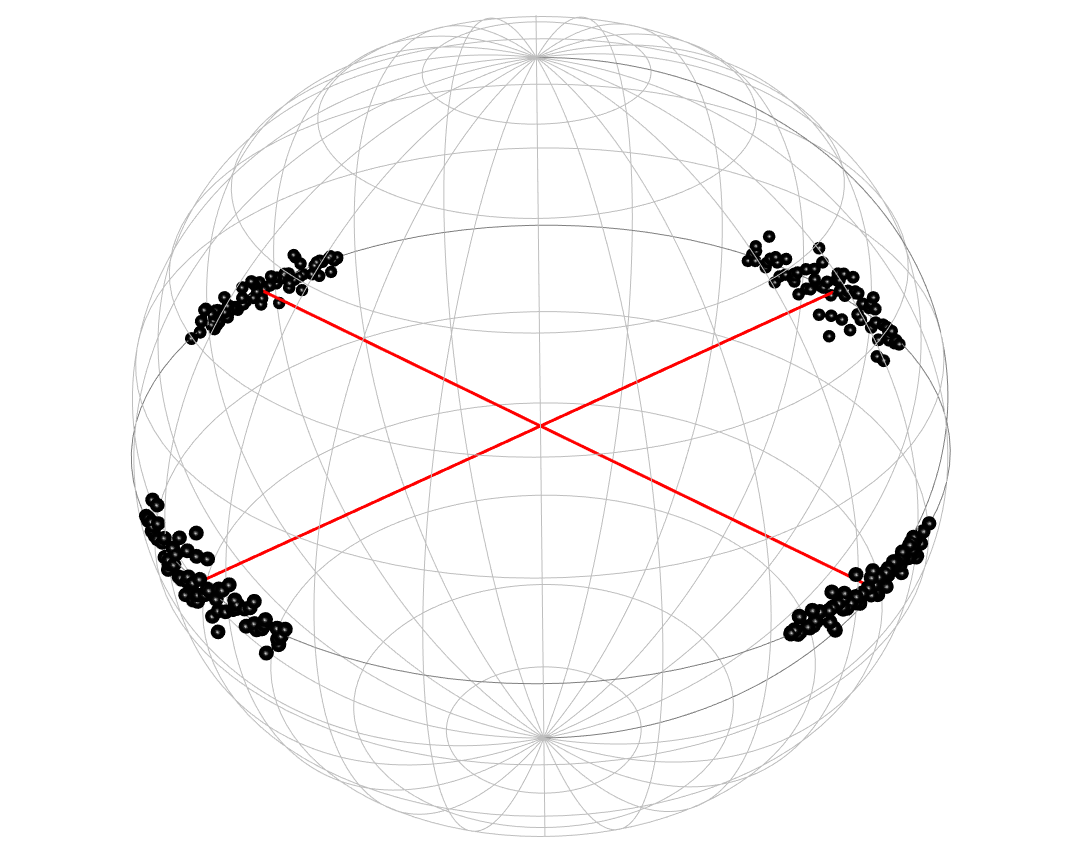}
\]
\vspace*{-0.5cm}
\caption{
  \revisedc{Direction clustering.}
  Left: all estimated directions.  Middle: sizes of all estimated directions proportional to weights.  Right: estimated directions with weights larger than a threshold. Red lines represents underlying true directions.}\label{fig:separate}
\end{figure}

Next we need a clustering strategy
to choose the number of clusters adaptively.
With the distance metric (\ref{eqn:dist2}), one can define dissimilarity matrix for a set of directions and make use of a generic clustering algorithm. Our choice is the Partition Around Medoids (PAM) \citep{Kaufman-Rousseeuw90} due to its simplicity.
Also, we apply the
average silhouette \citep{Rousseeuw87} to choose the number of clusters; see \suppref{Algorithm~S3} of the Supplemental Material.
The silhouette of a datum $i$ measures the strength of its membership to its cluster, as compared to the neighboring cluster.  Here, the neighboring cluster is the one, apart from cluster of datum $i$, that has the smallest average dissimilarity with datum $i$.  The corresponding silhouette is defined as
$(b_i-a_i)/(\max\{a_i,b_i\})$,
where $a_i$ and $b_i$ represent the average dissimilarities of datum $i$ with all other data in the same cluster and that with the neighboring cluster respectively.
The average silhouette of all data gives a measure of how good the clustering
is.  Thus we select the number of clusters via maximizing the average silhouette.
The detailed smoothing procedure is given in \suppref{Algorithm~S4}.


\subsection{Theoretical results}\label{sec:theory}
This subsection derives asymptotic properties of the proposed direction
smoothing estimator.
Note that, since the space of direction vectors has a non-Euclidean geometry and
so the theoretical framework is different from that of classical smoothing estimators.
Without loss of generality, suppose we observe $\bmv_1,\dots,\bmv_n \in \cM$
at spatial locations $\bms_1,
\dots, \bms_n$ respectively.
Let $\cS$ be the 3D unit sphere.
Then $\cM$ is the quotient space of $\cS$ with equivalence relation $\bmv \sim -\bmv$
for any $\bmv\in\cS$. This space is also identified with the so-called
real projective space $\mathbb{R} P^2$.

The theoretical results below were derived under \revisedb{the more convenient
  random design} where
$\bms_i$'s are independently and identically sampled from a distribution with
density $f_S$.}
  \dpaul{The below theorem (Theorem \ref{thm:con+norm}) remains valid even
    under a fixed, regular design setting, with the
    number of grid points increasing to infinity. In this case, in the statement of the asymptotic
    formulae and their proofs, the density function $f_S$ is replaced with a
    constant-valued function, representing a regular grid, with corresponding changes
    wherever derivatives of $f_S$ appear.}

  Given a spatial location $\bms_0$, our target is to estimate $\bmv_0$, namely
the diffusion direction at $\bms_0$,  \jpeng{which is defined as the minimizer of}
$
  \E \left\{d^{*2}(\bmV, \bmv) | \bmS=\bms_0 \right\}
$
\rwong{over $\bmv$,}
where ${d}^{*}(\bmu,\bmv)=\acos(|\bmu^\tp \bmv|)$.
\rwong{ Here $\bm{\mathrm{V}}$
is a random unit vector representing a random diffusion direction
and the expectation is taken over
$\bm{\mathrm{V}}$ conditional on $\bm{\mathrm{S}}=\bm{\mathrm{s}}_0$, where
$\bm{\mathrm{S}}$ represents the location of where
$\bm{\mathrm{V}}$ is observed.}
For simplicity, we assume $\bms_i\in\mathbb{R}$ and write it as $s_i$
thereafter. Thus, our estimator~(\ref{eqn:general_smooth}) at $s_0$ can be written as
\[
  \hat{\bmv}(s_0) = \argmin_{\bmv\in\cM} \sum^n_{i=1} K_h(s_i-s_0) d^{*2}(\bmv_i,\bmv),
\]
where \rwong{$n$ is the number of diffusion direction vectors} and
$K_h(\cdot)=K(\cdot/h)/h$. Here, with
slight notation abuse, $K(\cdot)$ represents a one dimensional kernel function throughout
the theoretical developments.

We first describe a working coordinate system.
For each $\bmp\in\cS$, one can endow a tangent space
$T_{\bmp} \cS=\{\bmv\in\mathbb{R}^3:\bmv^\tp \bmp=0\}$ with the metric tensor
$g_{\bmp} : T_{\bmp} \mathcal{V} \times T_{\bmp} \mathcal{V} \rightarrow \mathbb{R}$ defined as
$g_{\bmp}(\bmu_1, \bmu_2)= \bmu_1^\tp \bmu_2$.
Note that the tangent space is identified with $\mathbb{R}^2$.
The geodesics are great circles and the geodesic
distance is $\acos(\bmp_1^\tp \bmp_2)$, for any $\bmp_1, \bmp_2 \in \cS$. The corresponding exponential
map at $\bmp\in\cS$, $\Exp_{\bmp}:T_{\bmp}\cS\rightarrow\cS$, is given by
\begin{align*}
  \Exp_{\bmp}(\bm{0}) = \bmp \quad \mbox{and} \quad
  \Exp_{\bmp} (\bmu) = \cos(\|\bmu\|) \bmp + \frac{\sin(\|\bmu\|)}{\|\bmu\|}
  \bmu \quad \mbox{when} \quad \bmu\neq \bm{0},
\end{align*}
while the corresponding logarithm map at $\bmp\in\cS$, $\Log_{\bmp}:
\cS\backslash\{-\bmp\}\rightarrow T_{\bmp}\cS$, is
given by
\begin{align*}
  \Log_{\bmp}(\bmp) = \bm{0} \quad \mbox{and} \quad
  \Log_{\bmp}(\bmv) = \frac{\acos(\bmv^\tp \bmp)}{\sqrt{1-(\bmv^\tp \bmp)^2}}
  [ \bmv - (\bmv^\tp \bmp) \bmp ] \quad \mbox{when} \quad
  \bmv\neq \bmp.
\end{align*}
One can use the exponential map and the logarithm map to
define a coordinate system for the $\cS\backslash\{-\bmv_0\}$ in the following way.
Given $\bmv \in\cS$, we
define the logarithmic coordinate as
\[\omega_1 = \bme_1^\tp \Log_{\bmv_0} (\bmv) \quad \mbox{and} \quad \omega_2 = \bme_2^\tp \Log_{\bmv_0} (\bmv),
\]
where $\bme_1, \bme_2 \in T_{\bmv_0}\cS$ and $\{\bme_1, \bme_2\}$ forms an
orthonormal basis for $T_{\bmv_0}\cS$.
Write ${\phi}(\bmv) = (\omega_1, \omega_2)^\tp$.
In addition, we define
\[
  \rho_{\bmv_0}(\bmv) =\begin{cases}
    \mathrm{sign}(\bmv_0^\tp \bmv) \bmv & \bmv_0^\tp \bmv \neq 0\\
    \bmv & \bmv_0^\tp \bmv = 0
  \end{cases},
\]
\reviseda{which aligns $\bmv$ with $\bmv_0$, and
${d}(\bm{\omega},\bmtheta) = d^*({\phi}^{-1} (\bm{\omega}),{\phi}^{-1}
(\bmtheta))$ for $\bm{\omega}, \bmtheta\in \mathbb{R}^2$.
Note that for any $\bmv, \bmp\in\mathcal{V}$,
we have $d(\tilde{\phi}({\bmv}), \tilde{\phi}(\bmp)) = d^*(\bmv, \bmp)$
where $\tilde{\phi} = {\phi}\circ \rho_{\bmv_0}$.}
\revisedb{
  Here $\tilde{\phi}(\bmv)$ first aligns a direction $\bmv$ with
  the true diffusion direction $\bmv_0$ and then represents it by its
  logarithmic coordinate.
}

We now present the asymptotic results.
Now, write
$\bmtheta_i=\tilde{\phi}(\bmv_i)$
 for $i=1,\dots, n$, and $\psi(\bm{\omega},\bmtheta) =
d^2(\bm{\omega},\bmtheta)$. We have $\bmtheta_0 = \tilde{\phi}(\bmv_0)=\bm{0}$.
\rwong{Also, let $\bm{\psi}_j(\bm{\omega},\bmtheta)$ be the $j$-th order derivative of $\psi$ with
respect to $\bmtheta$ for $j=1,2$.}
Let $\brm{m}(s) = (m_1(s), m_2(s))^\tp = \E(\bmtheta_1|S_1=s)$ and $\bm{\Sigma}(s) =
[\Sigma_{jk}(s)]_{1\le j, k \le 2} =  \Var(\bmtheta_1 | S_1 =s)$. Also, denote
$\bm{\Psi}(s)= [\Psi_{jk}(s)]_{1\le j,k \le 2} = \E[\bm{\psi}_2(\bmtheta_1,
\bmtheta_0)|S_1=s]$.

Under the assumptions 1-10 laid out in \suppref{Section S5.2} of the Supplemental Material
\revisedc{which are all standard technical conditions (except for Assumption 1
  which is to ensure the representation of the geodesic distance as a function
  of the working coordinate system)},
we have the following theorem.
\begin{thm}\label{thm:con+norm}
  Let $M_n(\bmtheta) = \sum^n_{i=1} h K_h(S_i-s_0) d^2(\bmtheta_i, \bmtheta)$, and assume Assumptions 1-10 hold.

  \begin{enumerate}[(a)]
    \item There exists a sequence of solutions, $\hat{\bmtheta}_n(s_0)$, to
      $M_n^{(1)}(\bmtheta) = 0$,
  such that $\hat{\bmtheta}_n(s_0)$ converges in probability to $\bmtheta_0$.
\item $\hat{\bmtheta}_n$ is asymptotically normal:
  \[
    \sqrt{nh}\left\{ (\hat{\bmtheta}_n-\bmtheta_0) - h^2 \bm{\eta}\right\} \implies
    \mathcal{N}_2(\bm{0},\bm{\Omega}),
  \]
  where
  \[
    \bm{\eta} = 2\int x^2 K(x) dx \bm{\Psi}^{-1}(s_0) \left\{
    \frac{\fs^{(1)}(s_0)}{\fs(s_0)}m^{(1)}(s_0) +
    \frac{1}{2} m^{(2)}(s_0)\right\}
  \]
  and
  \[
    \bm{\Omega} = 4 \int K^2(x) dx \bm{\Psi}^{-1}(s_0) \bm{\Sigma}(s_0).
  \]
\end{enumerate}
\end{thm}
The proof of the Theorem~\ref{thm:con+norm} can be found in \revisedb{\suppref{Section~S5.2}} of the Supplemental Material.

\section{Fiber tracking}
\label{sec:track}
For dMRI, fiber tractography can be \revisedc{classified as} deterministic and probabilistic methods.  Deterministic methods \citep[e.g.][]{Mori-Crain-Chacko99, Weinstein-Kindlmann-Lundberg99, Mori-Zijl02} track fiber bundles by utilizing the principal eigenvectors of tensors, while probabilistic methods \citep[e.g.][]{Koch-Norris-Hund-Georgiadis02,Parker-Alexander03,Friman-Farneback-Westin06} use the probability density of diffusion orientations.
Most \revisedc{deterministic} methods \revisedd{assume one} single diffusion tensor in each voxel, and hence \revisedd{are unable} to handle voxels with crossing fibers.  In view of this, this section develops a \revisedc{deterministic} tracking algorithm that allows for multiple or no principal diffusion directions in a voxel.

\begin{figure}[htpb]
  \[
  \includegraphics[height=3.5cm]{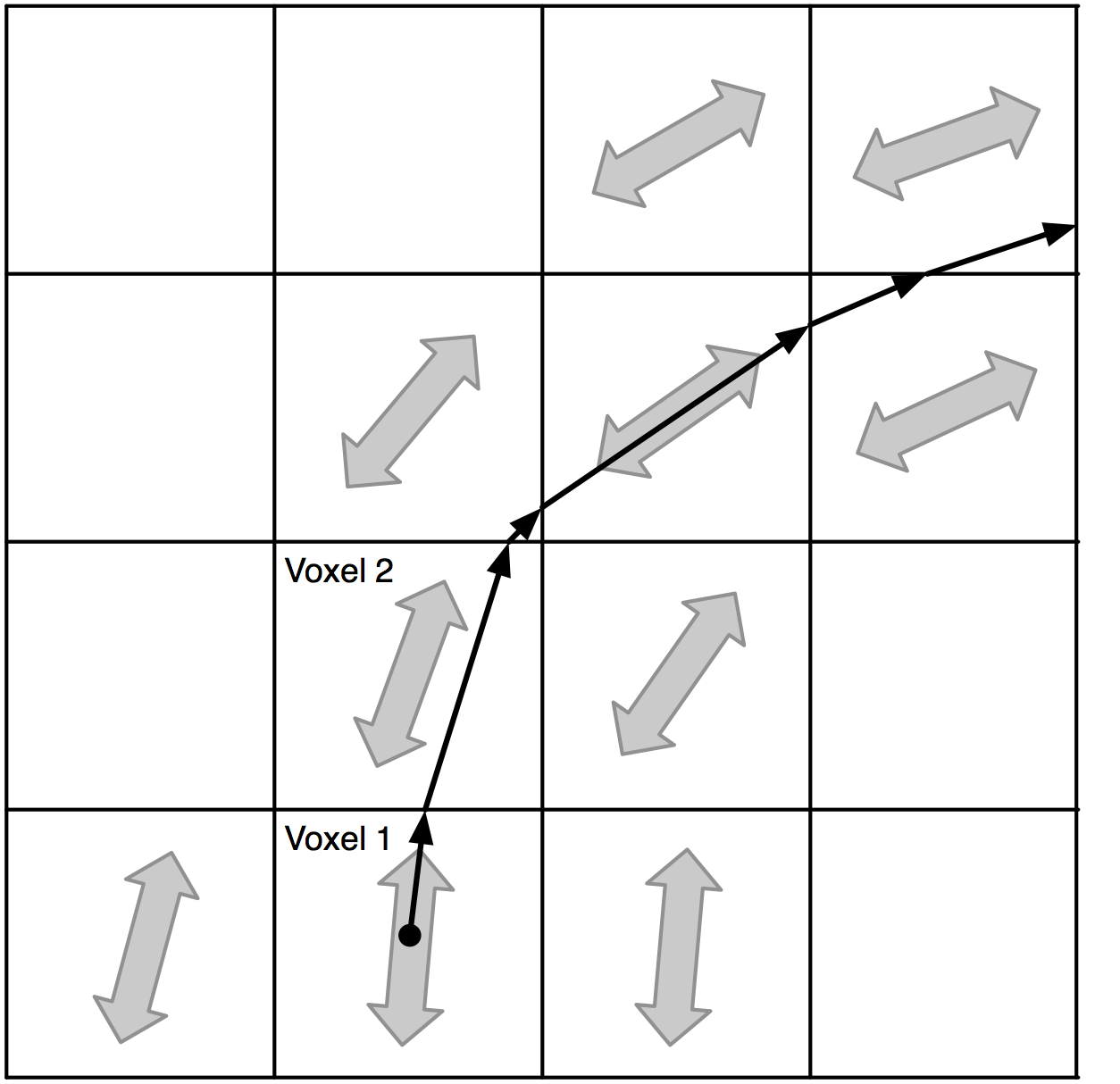}
  \hspace*{0.2cm}
  \includegraphics[height=3.5cm]{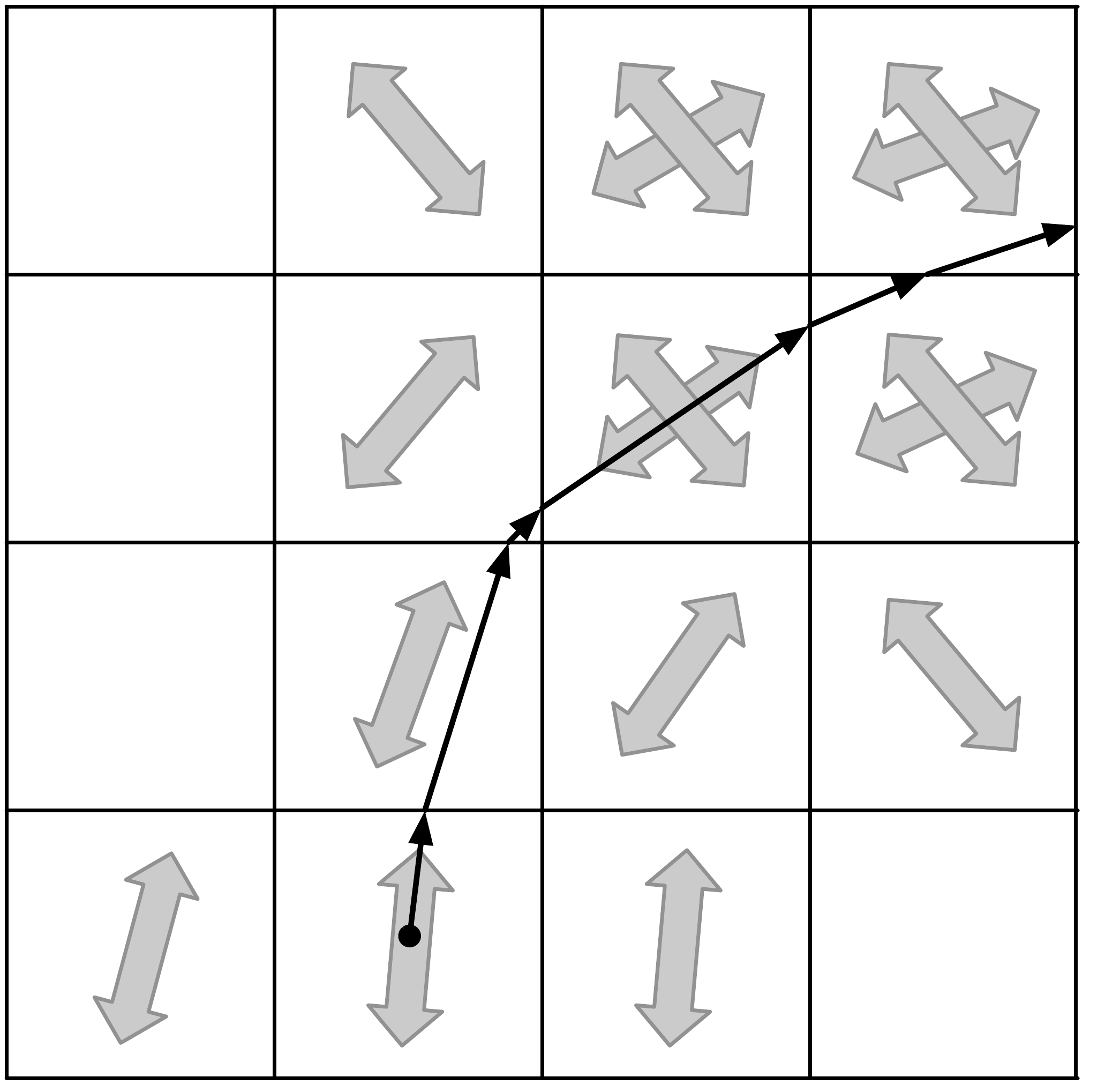}
  \hspace*{0.2cm}
  \includegraphics[height=3.5cm]{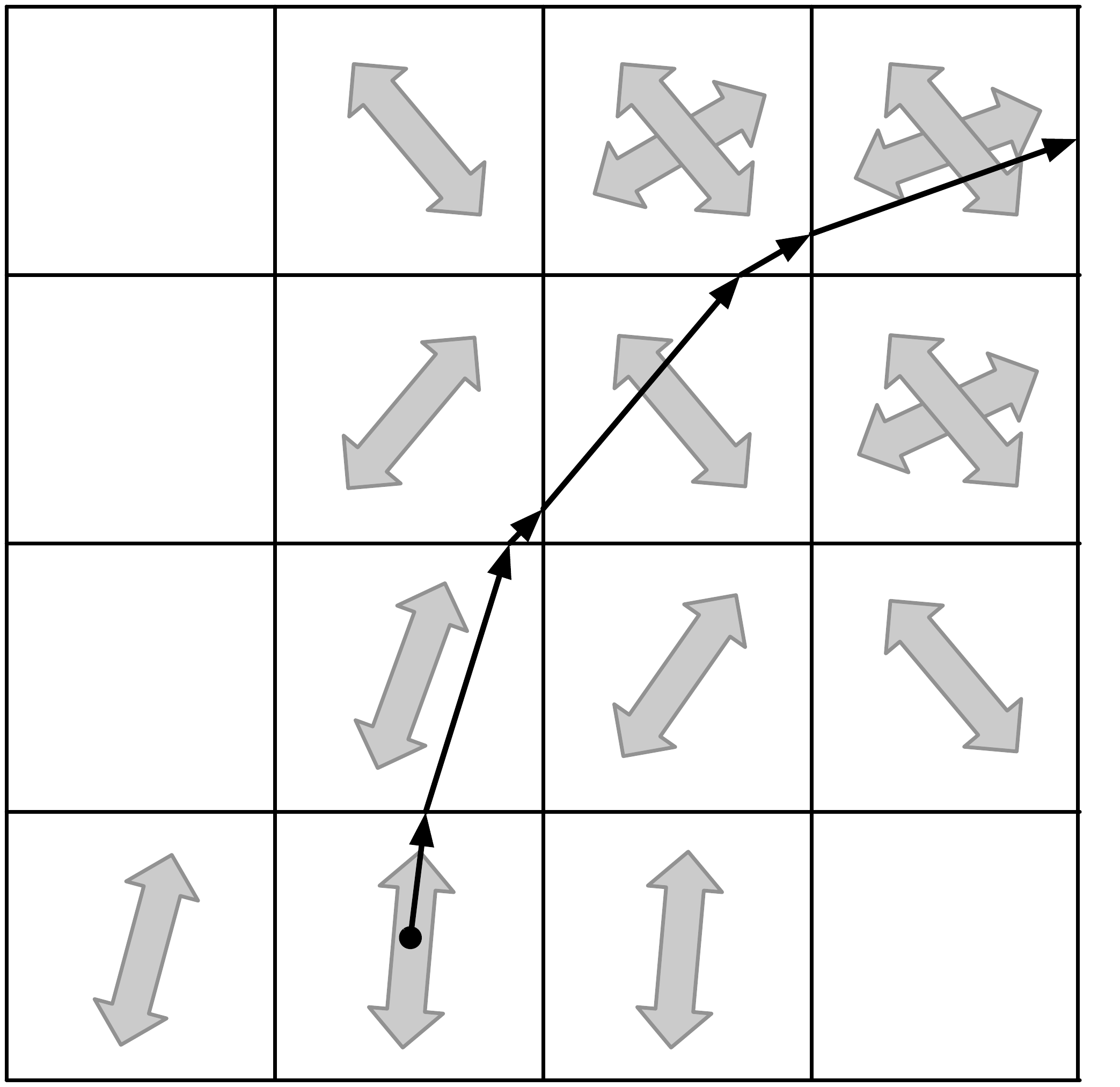}
  \]
  \caption{Left: Demonstration of the proposed algorithm in single fiber region. Middle:
    Demonstration of the proposed algorithm in crossing fiber region. Right:
    Demonstration of the proposed algorithm in case of absence of viable
    directions.}\label{fig:tract}
\end{figure}

The proposed algorithm can be seen as a generalization of the popular Fiber Assignment by Continuous Tracking (FACT) \citep{Mori-Crain-Chacko99} algorithm.  A brief description of FACT is as \revisedc{follows.}
Tracking starts at the center of a voxel (Voxel~1 in Figure~\ref{fig:tract} left panel) and continues in the direction of the estimated diffusion direction.  When it enters the next voxel (Voxel~2 in Figure~\ref{fig:tract} left panel), the track changes its direction to align with the new diffusion direction and so on.  This tracking rule may produce many short and fragmented fiber tracts due to either a wrongfully identified isotropic voxel or spurious directions which go nowhere.  In addition, it cannot determine which direction to follow in case there are multiple directions in a voxel, which happens in crossing fiber regions.

To address these issues, we modify the above procedure in the following manner.  Given a current diffusion direction (we refer to the corresponding voxel as the current voxel), the voxel that it points to (we refer to this voxel as the destination voxel) may have (i) at least one direction; (ii) no direction (i.e., isotropic).  In case (i), we will first identify the direction with the smallest angular difference with
the current direction. If its separation angle is smaller than a pre-specified
threshold (e.g., $\pi/6$), we enter the destination voxel and tracking
will go on along this direction. See Figure~\ref{fig:tract} (Middle).
On the other hand, if the separation angle is
greater than the threshold, or case (ii) happens, we deem that the destination
voxel does not have a viable direction. In this case, tracking will go along
the current direction if it finds a viable direction within a pre-specified
number of voxels.  The number of voxels that are allowed to be skipped is
set to be $1$ in our numerical illustrations.
See Figure~\ref{fig:tract} (Right). On the other hand,
the tracking stops at the current voxel if no viable directions within a
pre-specified number of voxels can be found.
The detailed tracking algorithm is described in \suppref{Algorithm~S5} in the
Supplemental Material.

As for the choice of starting voxels, also known as seeds, there are two common strategies. One can choose seeds based on tracts of interest and start the tracking from a region of interest (ROI). This approach is based on knowledge on ROI and may not give a full picture of the tracts of interest if there are diverging branches. The other approach is \revisedc{the} brute-force approach, where tracking starts from every voxel. It usually leads to a more comprehensive picture of tracts at a higher computational cost.  The proposed algorithm can be coupled with either strategy.

Combining the voxel-wise estimation method in Section~\ref{sec:voxelwise} and the direction smoothing procedure in Section~\ref{sec:smoothing} gives the proposed DiST method.

%
\section{Simulation \revisedc{study}}\label{sec:sim1}
\tlee{Extensive simulation experiments have been conducted to evaluate the practical performances of DiST.  They are reported in \suppref{Section~S6} of the Supplemental Material.  Overall, the DiST method provided highly promising results.}

\section{Real data application}\label{sec:real}
In this section, we apply the proposed methodology to a real dMRI data set,
which was obtained from the
Alzheimer's
Disease Neuroimaging Initiative (ADNI) database (www.loni.ucla.edu/ADNI).
The primary
goal of ADNI has been to test whether serial MRI,
positron emission tomography (PET), other biological markers, and clinical and
neuropsychological assessment can be combined to measure the progression of
mild cognitive impairment (MCI) and onset of Alzheimer's disease (AD).
In the following, we use an eddy-current-corrected ADNI data set of a normal
subject for illustration of our technique.

This data set contains 41 distinct gradient directions with
$b$-value set as $1000s/mm^2$. In addition, there are 5 $b0$ images (corresponding to
$b=0$), forming
in total 46 measurements for each of the $256\times256\times59$ voxels.
To implement our technique, we require estimates of $S_0(\bms)$'s and $\sigma$.
We first estimate
$S_0(\bms)$ and $\sigma(\bms)$ for each voxel by ML estimation based on the
5 $b0$ images.
Then we fix
$\sigma$  as the median of estimated $\sigma(\bms)$'s for voxel-wise estimation
of the diffusion directions.
Since the original $256\times256\times59$ voxels contain volume outside the brain,
we only take median over a human-chosen set of $81\times81\times20$ voxels.
The estimated $\sigma$ is 56.9.

\begin{figure}[htpb]
  \centering
  \includegraphics[height=4.1cm]{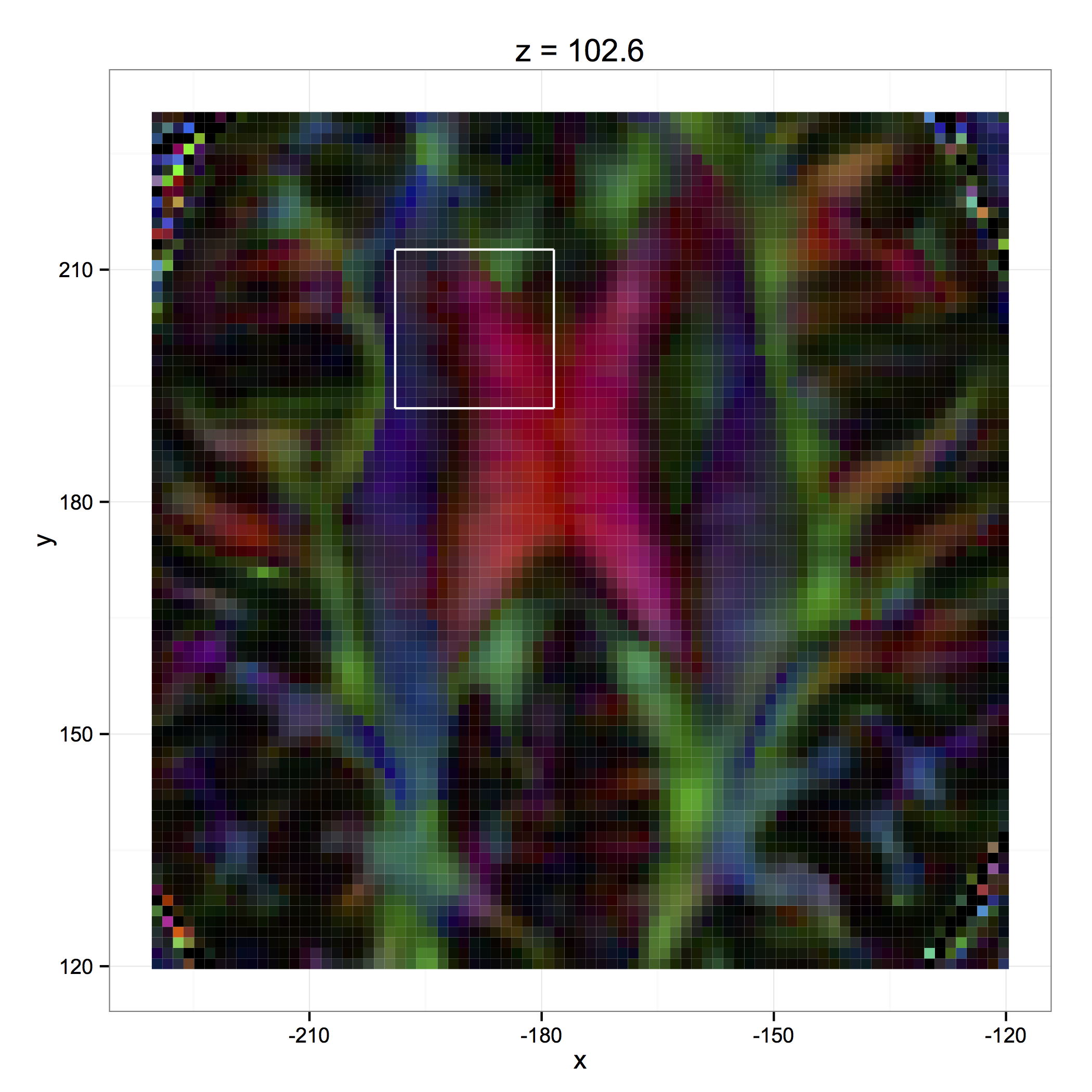}
  \includegraphics[height=4.1cm]{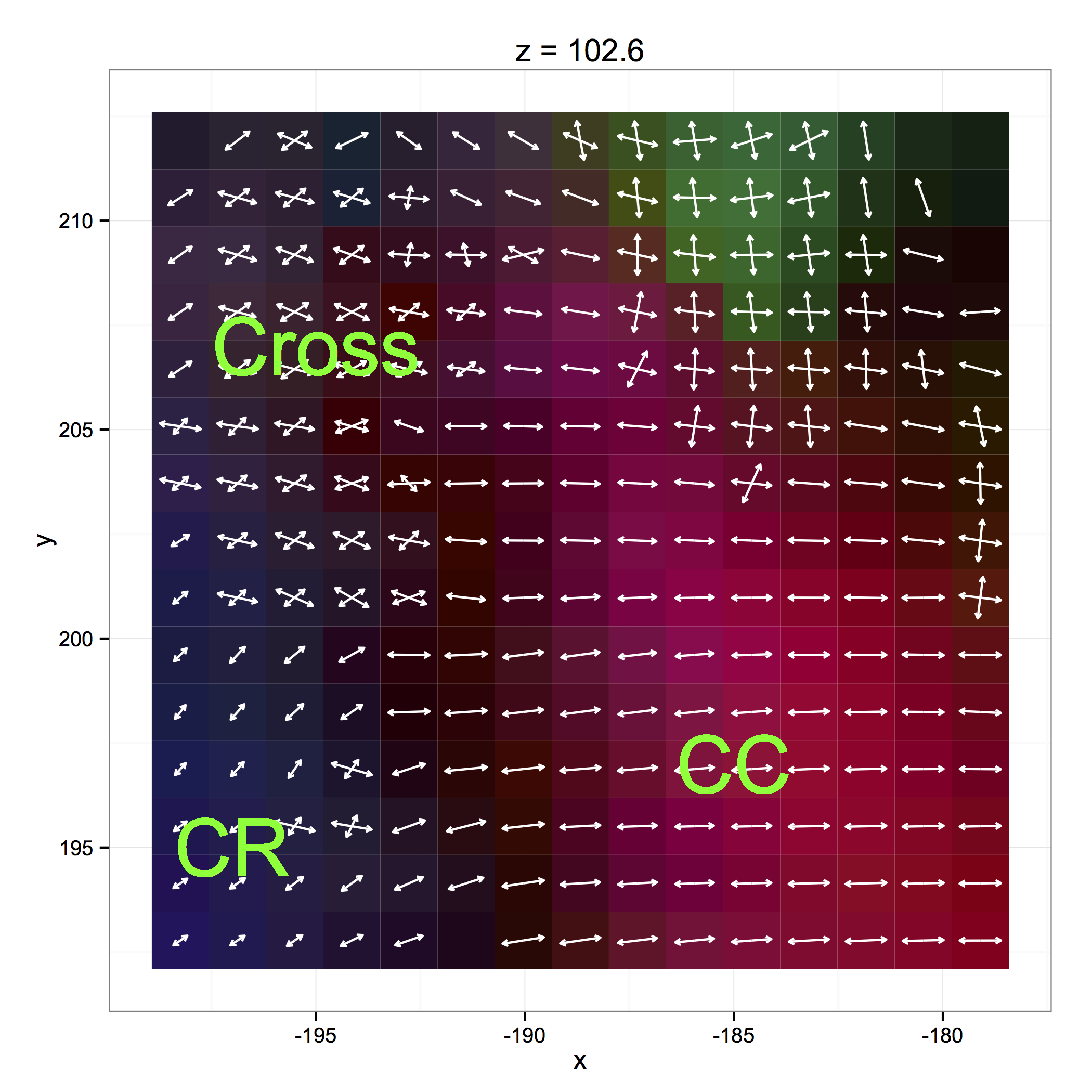}
  \includegraphics[height=4.1cm]{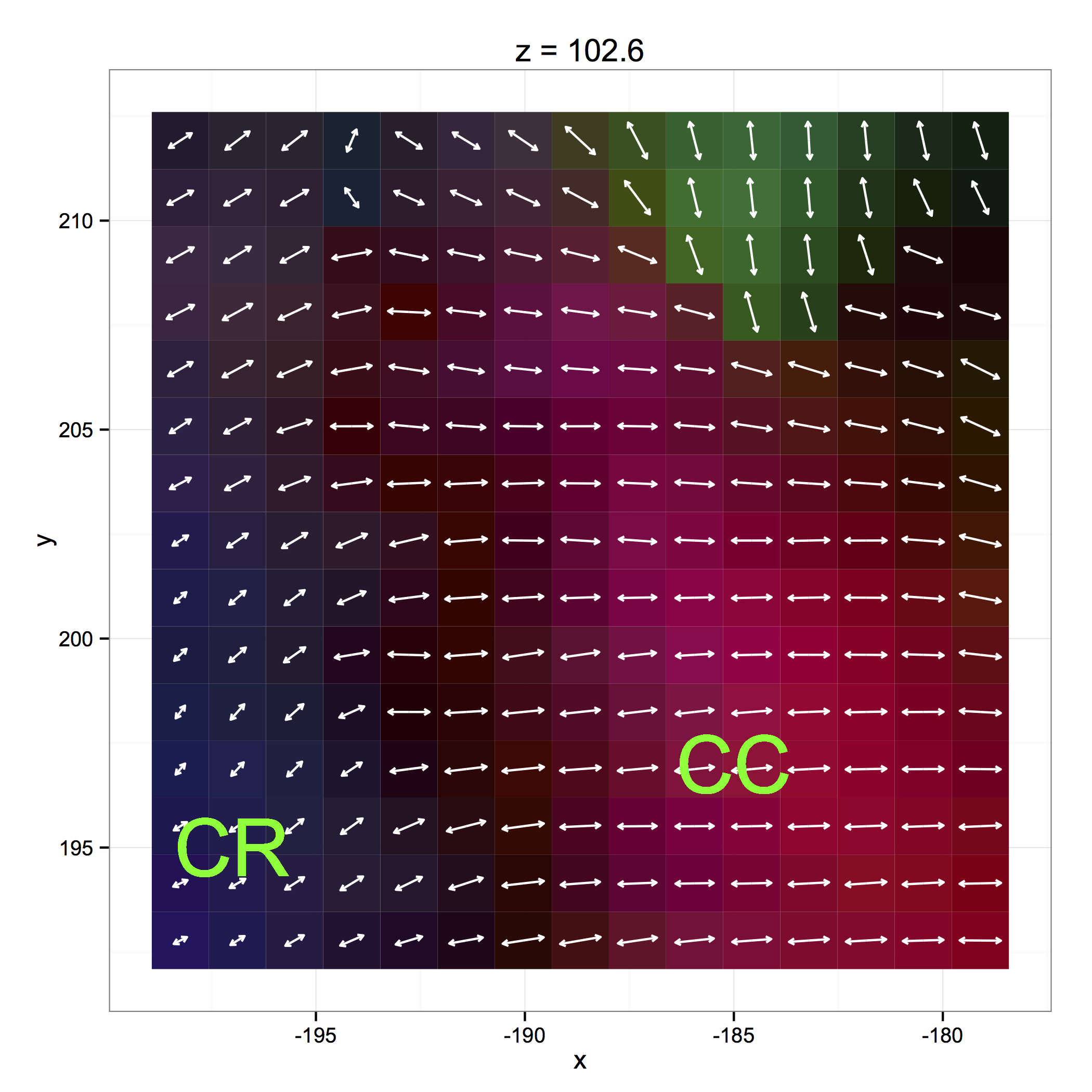}
  \caption{\revisedb{Left: the fiber orientation color map (based on the single tensor
    model). The focused region is indicated by white rectangular box. Middle \revisedc{(from \textsf{DiST-mcv})}
    and right \revisedc{(from single tensor model)}: The projection of fiber
    directions to the $xy$-plane
    at $z=102.6$
    for illustration of crossing fibers.
    \rwong{(The five $xy$-planes that we focus on have reference values $z=99.9, 102.6, 105.3, 108, 110.7$ from bottom to top.)}
    The plot also shows the location of corpus callosum
    (CC), corona radiata (CR) and crossing region (Cross).
  The fiber orientation color map is overlaid as the background.}
}\label{fig:project}
\end{figure}

In this analysis, we focus on a subset of voxels ($15\times15\times5$), which contains the intersection of corpus callosum (CC) and corona radiata (CR).  This region is known to contain significant fiber crossing \citep{Wiegell-Larsson-Wedeen00}.
\revisedb{See Figure~\ref{fig:project} (Left) for a fiber orientation color map of one of the five $xy$-planes.
Within the whole focused region, $S_0(\bms)$'s have mean 1860.1 and standard deviation 522.7.}

We then apply voxel-wise estimation to individual voxels followed by the \textsf{DiST-mcv}
procedure.
Distributions of the estimated number of diffusion directions are summarized in Table
\ref{tab:real:hatJ}.
For comparison purposes, we also fit the single tensor model with the commonly
used regression estimator \citep[e.g.,][]{Mori07}.

\begin{figure}[htpb]
  \vspace{-0.5cm}
  \[
    \includegraphics[height=5cm]{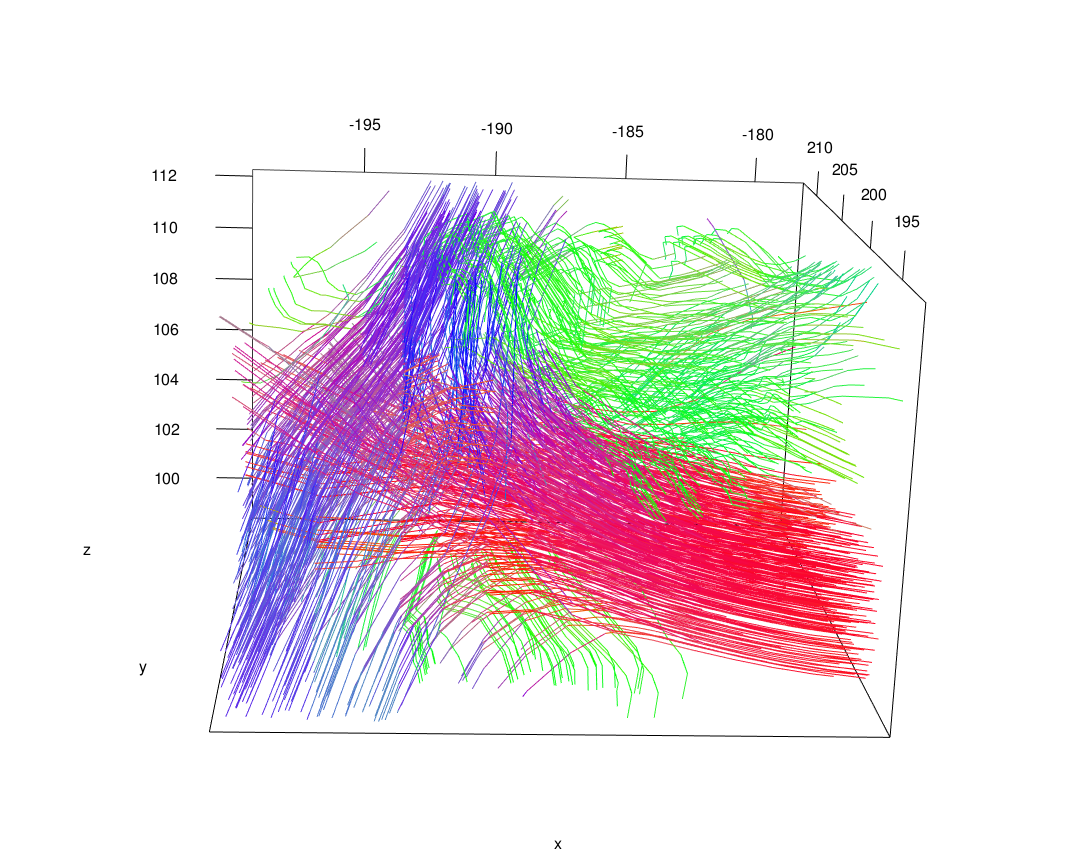}
    \hspace*{-0.5cm}
    \includegraphics[height=5cm]{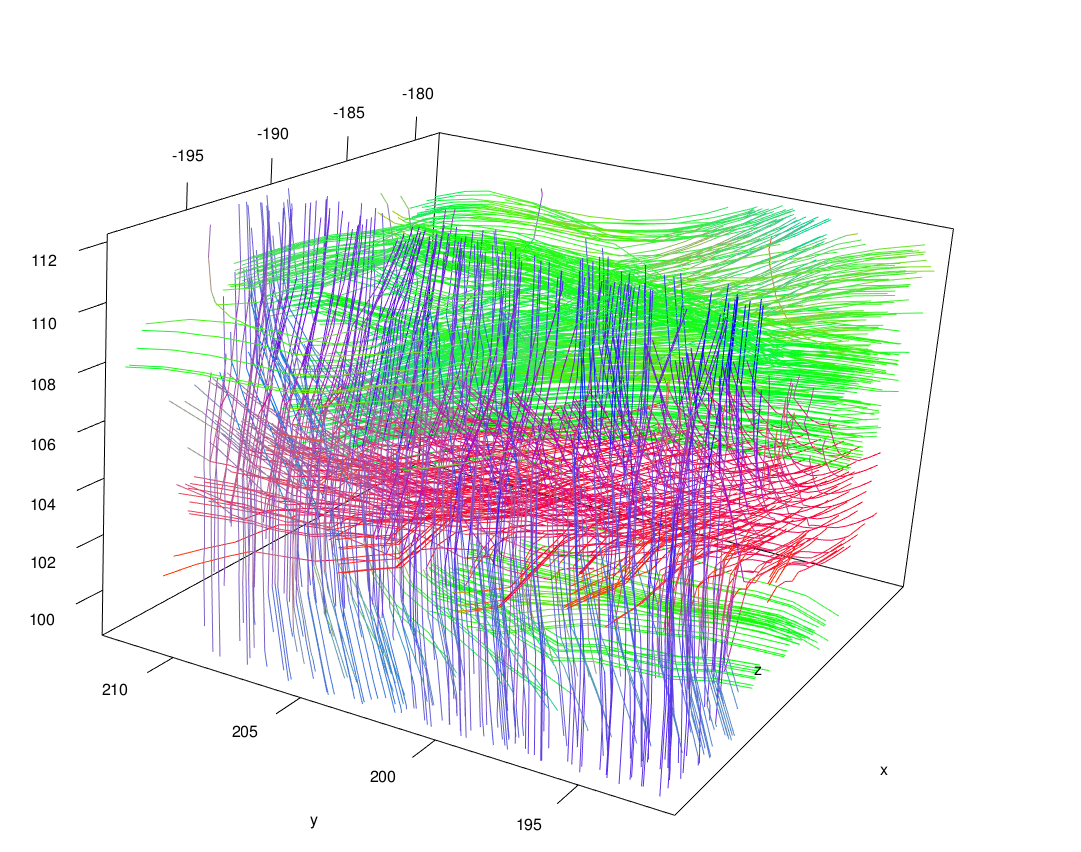}
  \]
  \vspace{-0.5cm}
  \[
    \includegraphics[height=5cm]{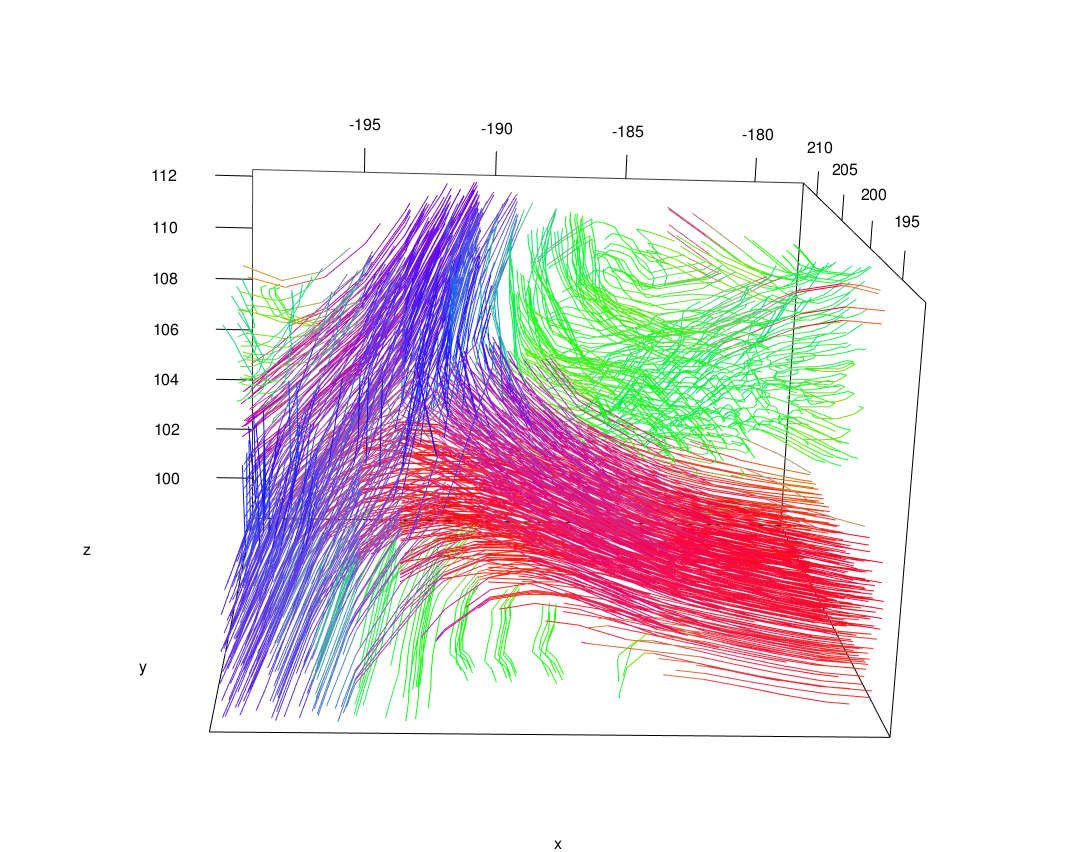}
    \hspace*{-0.5cm}
    \includegraphics[height=5cm]{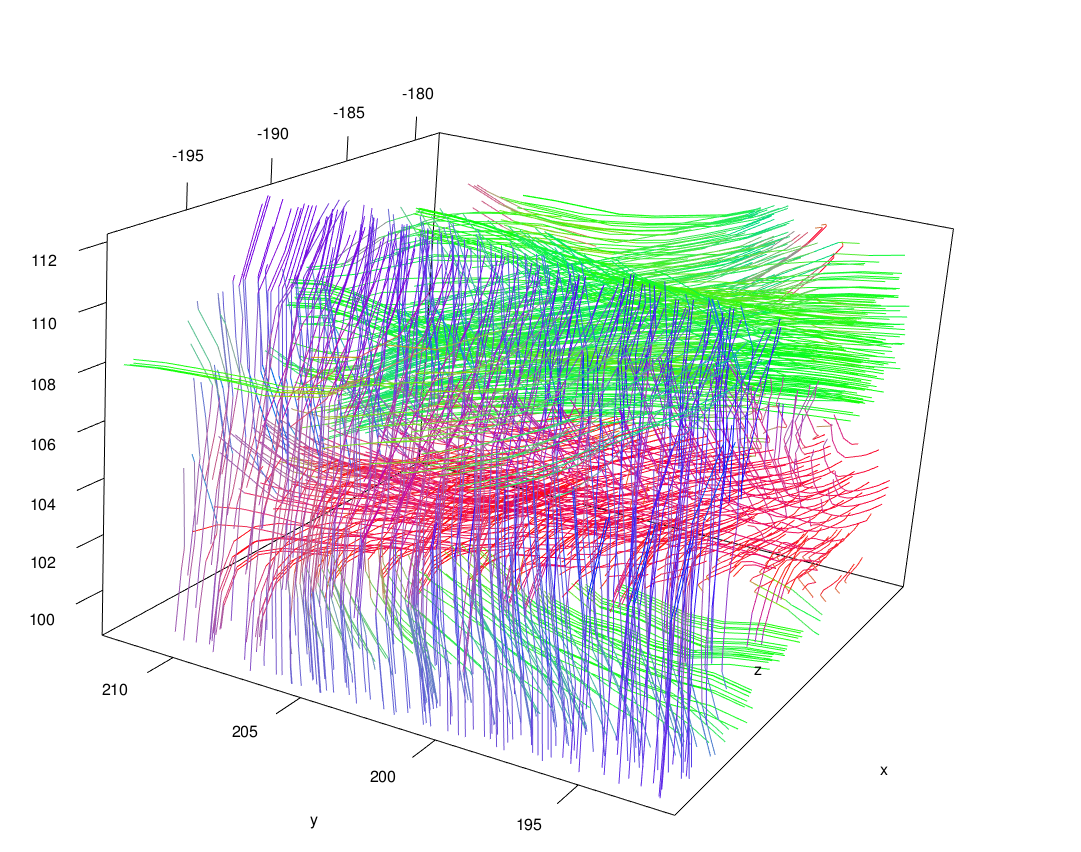}
  \]
  \caption{Top: The longest 900 tracks using \textsf{DiST-mcv}.
    Bottom: \revisedc{The longest 900 tracks using the} single tensor model.
    The left and right figures correspond to different viewing angles.
  }
  \label{fig:corpus_track_dsmooth_level}
\end{figure}

The tracking results are produced by applying the proposed tracking algorithm to the estimated diffusion directions from \textsf{DiST-mcv},
\rwong{which represents the DiST procedure with $h$ chosen by the median cross-validation score
(see Sections S3 and S6 of the Supplemental Material),}
and those from the single tensor model estimation.
\revisedb{For visualization purposes, we present the longest 900 tracts in Figure~\ref{fig:corpus_track_dsmooth_level}.}
From anatomy, the CC has a mediolateral direction
while the CR has a superoinferior orientation. They are clearly shown in both
tracking results.
In these figures, reconstructed fiber tracts are colored by a RGB color model
with red for left-right, green for anteroposterior, and blue for superior-inferior.
Thus, one can easily locate the CC and the CR as the red fiber bundle and the
blue fiber bundle respectively.
Tracking result based on \textsf{DiST-mcv} shows clear crossing between
mediolateral fiber and the superoinferior fiber (in the figure, the crossing of
red and blue fiber tracts).
From neuroanatomic
atlases and previous studies, \citet{Wiegell-Larsson-Wedeen00} conclude that
there are several fiber populations with crossing structure in this
conjunction region of CC and CR,
which matches with the tracking based on \textsf{DiST-mcv}.
However, the single tensor model estimation can only
reconstruct one major diffusion direction in each voxel and thus
the corresponding tracking result does not show crossing structure.
Instead, the CC (red fiber bundle) is blocked by the CR (blue fiber bundle) and this
leads to either termination of the CC fiber tracts or significant merging of the CC and
the CR fiber tracts instead of the known crossing structure.
To give further illustration,
Figure~\ref{fig:project} shows the locations of the CC, the CR and the region of
crossing fibers (Cross).
One can see that estimated directions based on \textsf{DiST-mcv} reproduces the crossing
fiber structures between the CC and the CR, while the result based on single tensor
model tends to connect the CC and the CR fibers.

Moreover, the green fiber on top of the CC represents the cingulum bundle.
Both fiber tracking based on {DiST} and single fiber model
produce clear and sensible reconstruction of cingulum bundle.
All these features match with neuroanatomic atlases and provide
a good demonstration of our proposed method.

\jpeng{As shown by Figures~\ref{fig:project} and~\ref{fig:corpus_track_dsmooth_level}, when comparing with the results obtained by the single tensor model, {DiST} produces more biologically sensible and interpretable tracking results.  This provides more reliable information on brain connectivity and in turn could lead to better understanding of neuro-degenerative diseases such as Alzheimer's disease and autism as well as  better detection of brain abnormality, such as deformation and neuron loss in white matter regions.}

\linsps
\begin{table}[htpb]
  \centering
  \caption{Number of voxels with different estimated number of diffusion
    directions.}\label{tab:real:hatJ}
  \vspace{0.3cm}
  {\small
    \begin{tabular}{c|ccccc|c}
      \hline\hline
      & \multicolumn{5}{c|}{Number of diffusion directions}\\
      & 0 & 1 & 2 & 3 & 4 & \rwong{total} \\
      \hline\hline
      Voxel-wise estimation & 37 & 476 & 589 & 23 & 0 & \rwong{1125} \\
      Smoothing &  37  & 476 & 593 & 19 & 0 & \rwong{1125}\\
      \hline\hline
    \end{tabular}
  }
\end{table}
\linsp

\section{Discussion}
\label{sec:discuss}
Using tensor estimation to resolve crossing fiber can be problematic,
\revisedc{due to the inability of estimating multiple diffusion directions by the single tensor model and}
the non-identifiability issue in multi-tensor model.  In this paper, we take a
different route by focusing on the estimation of diffusion directions rather
than the diffusion tensors.  We develop the corresponding direction smoothing
procedure and fiber tracking strategy, together called DiST, along this route.
Our technique gives promising empirical results in both simulation study
\rwong{(see Section S6 of the Supplemental Material)}
and real data analysis.

The procedure we presented works well  even with moderate number of gradient directions (a few tens), as long as the number of distinct crossing fibers within a voxel is not \tlee{larger than three}. With HARDI data, which can have up to a couple of hundreds gradient directions, rather than modeling the direction distribution within a tensor framework, we can estimate the fiber orientation distribution nonparametrically \citep{Tuch04, Descoteaux-Angelino-Fitzgibbons07}.

\ignore{In that case, we can potentially extend the fiber tracking procedure presented here by adopting a probabilistic approach in which the directions for moving from one voxel to another are sampled from the fiber orientation distribution. Such a probabilistic fiber tracking  has the additional advantage of giving a  measure of uncertainty of the fiber tracts extracted from the data. This is a topic of future research.}

\jpeng{Applying {DiST} to multiple images from ADNI (either
  from the same subject over time or from multiple subjects) and then relating
  the tracking results with clinical outcomes such as cognitive measures would
  provide valuable information about the role of white matter connectivity in
  initiation and progression of Alzheimer's disease and dementia. Although this
  is an important direction of research, it is beyond the scope of this paper
  which focuses on developing a statistical procedure to denoise dMRI data and
  to provide better tracking results. We plan to explore more sophisticated
  applications of the proposed procedure in our future research.}

\section*{Acknowledgement}
Data collection and sharing for this project was funded by the Alzheimer's
Disease Neuroimaging Initiative (ADNI) (National Institutes of Health Grant U01
AG024904). ADNI is funded by the National Institute on Aging, the National
Institute of Biomedical Imaging and Bioengineering, and through generous
contributions from the following: Abbott, AstraZeneca AB, Bayer Schering Pharma
AG, Bristol-Myers Squibb, Eisai Global Clinical Development, Elan Corporation,
Genentech, GE Healthcare, GlaxoSmithKline, Innogenetics, Johnson and Johnson,
Eli Lilly and Co., Medpace, Inc., Merck and Co., Inc., Novartis AG, Pfizer Inc,
F. Hoffman-La Roche, Schering-Plough, Synarc, Inc., as well as non-profit
partners the Alzheimer's Association and Alzheimer's Drug Discovery Foundation,
with participation from the U.S. Food and Drug Administration. Private sector
contributions to ADNI are facilitated by the Foundation for the National
Institutes of Health (www.fnih.org). The grantee organization is the Northern
California Institute for Research and Education, and the study is coordinated
by the Alzheimer's Disease Cooperative Study at the University of California,
San Diego. ADNI data are disseminated by the Laboratory for Neuro Imaging at
the University of California, Los Angeles. This research was also supported by
NIH grants P30 AG010129, K01 AG030514, and the Dana Foundation.
The authors would like to thank Professor Owen Carmichael for making available
the data and his valuable comments.



\bibliographystyle{imsart-nameyear}
\bibliography{raywongref}

\begin{thebibliography}{39}

\bibitem[\protect\citeauthoryear{Arsigny {\it
  et~al.}}{2006}]{Arsigny-Fillard-Pennec06}
\begin{barticle}[author]
\bauthor{\bsnm{Arsigny},~\bfnm{Vincent}\binits{V.}},
  \bauthor{\bsnm{Fillard},~\bfnm{Pierre}\binits{P.}},
  \bauthor{\bsnm{Pennec},~\bfnm{Xavier}\binits{X.}} \AND
  \bauthor{\bsnm{Ayache},~\bfnm{Nicholas}\binits{N.}}
(\byear{2006}).
\btitle{Log-Euclidean metrics for fast and simple calculus on diffusion
  tensors}.
\bjournal{Magnetic resonance in medicine}
\bvolume{56}
\bpages{411--421}.
\end{barticle}
\endbibitem

\bibitem[\protect\citeauthoryear{Bammer {\it
  et~al.}}{2009}]{Bammer-Holdsworth-Veldhuis09}
\begin{barticle}[author]
\bauthor{\bsnm{Bammer},~\bfnm{Roland}\binits{R.}},
  \bauthor{\bsnm{Holdsworth},~\bfnm{Samantha~J}\binits{S.~J.}},
  \bauthor{\bsnm{Veldhuis},~\bfnm{Wouter~B}\binits{W.~B.}} \AND
  \bauthor{\bsnm{Skare},~\bfnm{Stefan~T}\binits{S.~T.}}
(\byear{2009}).
\btitle{New methods in diffusion-weighted and diffusion tensor imaging}.
\bjournal{Magnetic resonance imaging clinics of North America}
\bvolume{17}
\bpages{175--204}.
\end{barticle}
\endbibitem

\bibitem[\protect\citeauthoryear{Basser {\it
  et~al.}}{2000}]{Basser-Pajevic-Pierpaoli00}
\begin{barticle}[author]
\bauthor{\bsnm{Basser},~\bfnm{Peter~J}\binits{P.~J.}},
  \bauthor{\bsnm{Pajevic},~\bfnm{Sinisa}\binits{S.}},
  \bauthor{\bsnm{Pierpaoli},~\bfnm{Carlo}\binits{C.}},
  \bauthor{\bsnm{Duda},~\bfnm{Jeffrey}\binits{J.}} \AND
  \bauthor{\bsnm{Aldroubi},~\bfnm{Akram}\binits{A.}}
(\byear{2000}).
\btitle{In vivo fiber tractography using DT-MRI data}.
\bjournal{Magnetic Resonance in Medicine}
\bvolume{44}
\bpages{625--632}.
\end{barticle}
\endbibitem

\bibitem[\protect\citeauthoryear{Beaulieu}{2002}]{Beaulieu02}
\begin{barticle}[author]
\bauthor{\bsnm{Beaulieu},~\bfnm{Christian}\binits{C.}}
(\byear{2002}).
\btitle{The basis of anisotropic water diffusion in the nervous system -- a
  technical review}.
\bjournal{NMR in Biomedicine}
\bvolume{15}
\bpages{435--455}.
\end{barticle}
\endbibitem

\bibitem[\protect\citeauthoryear{Behrens {\it
  et~al.}}{2003}]{Behrens-Woolrich-Jenkinson03}
\begin{barticle}[author]
\bauthor{\bsnm{Behrens},~\bfnm{TEJ}\binits{T.}},
  \bauthor{\bsnm{Woolrich},~\bfnm{MW}\binits{M.}},
  \bauthor{\bsnm{Jenkinson},~\bfnm{M}\binits{M.}},
  \bauthor{\bsnm{Johansen-Berg},~\bfnm{H}\binits{H.}},
  \bauthor{\bsnm{Nunes},~\bfnm{RG}\binits{R.}},
  \bauthor{\bsnm{Clare},~\bfnm{S}\binits{S.}},
  \bauthor{\bsnm{Matthews},~\bfnm{PM}\binits{P.}},
  \bauthor{\bsnm{Brady},~\bfnm{JM}\binits{J.}} \AND
  \bauthor{\bsnm{Smith},~\bfnm{SM}\binits{S.}}
(\byear{2003}).
\btitle{Characterization and propagation of uncertainty in diffusion-weighted
  MR imaging}.
\bjournal{Magnetic Resonance in Medicine}
\bvolume{50}
\bpages{1077--1088}.
\end{barticle}
\endbibitem

\bibitem[\protect\citeauthoryear{Behrens {\it
  et~al.}}{2007}]{Behrens-Berg-Jbabdi07}
\begin{barticle}[author]
\bauthor{\bsnm{Behrens},~\bfnm{TEJ}\binits{T.}},
  \bauthor{\bsnm{Berg},~\bfnm{H~Johansen}\binits{H.~J.}},
  \bauthor{\bsnm{Jbabdi},~\bfnm{S}\binits{S.}},
  \bauthor{\bsnm{Rushworth},~\bfnm{MFS}\binits{M.}} \AND
  \bauthor{\bsnm{Woolrich},~\bfnm{MW}\binits{M.}}
(\byear{2007}).
\btitle{Probabilistic diffusion tractography with multiple fibre orientations:
  What can we gain?}
\bjournal{Neuroimage}
\bvolume{34}
\bpages{144--155}.
\end{barticle}
\endbibitem

\bibitem[\protect\citeauthoryear{Carmichael {\it
  et~al.}}{2013}]{Carmichael-Chen-Paul13}
\begin{barticle}[author]
\bauthor{\bsnm{Carmichael},~\bfnm{Owen}\binits{O.}},
  \bauthor{\bsnm{Chen},~\bfnm{Jun}\binits{J.}},
  \bauthor{\bsnm{Paul},~\bfnm{Debashis}\binits{D.}} \AND
  \bauthor{\bsnm{Peng},~\bfnm{Jie}\binits{J.}}
(\byear{2013}).
\btitle{Diffusion tensor smoothing through weighted Karcher means}.
\bjournal{Electronic Journal of Statistics}
\bvolume{7}
\bpages{1913--1956}.
\end{barticle}
\endbibitem

\bibitem[\protect\citeauthoryear{Chanraud {\it
  et~al.}}{2010}]{Chanraud-Zahr-Sullivan10}
\begin{barticle}[author]
\bauthor{\bsnm{Chanraud},~\bfnm{Sandra}\binits{S.}},
  \bauthor{\bsnm{Zahr},~\bfnm{Natalie}\binits{N.}},
  \bauthor{\bsnm{Sullivan},~\bfnm{Edith~V}\binits{E.~V.}} \AND
  \bauthor{\bsnm{Pfefferbaum},~\bfnm{Adolf}\binits{A.}}
(\byear{2010}).
\btitle{MR diffusion tensor imaging: a window into white matter integrity of
  the working brain}.
\bjournal{Neuropsychology review}
\bvolume{20}
\bpages{209--225}.
\end{barticle}
\endbibitem

\bibitem[\protect\citeauthoryear{Descoteaux {\it
  et~al.}}{2007}]{Descoteaux-Angelino-Fitzgibbons07}
\begin{barticle}[author]
\bauthor{\bsnm{Descoteaux},~\bfnm{Maxime}\binits{M.}},
  \bauthor{\bsnm{Angelino},~\bfnm{Elaine}\binits{E.}},
  \bauthor{\bsnm{Fitzgibbons},~\bfnm{Shaun}\binits{S.}} \AND
  \bauthor{\bsnm{Deriche},~\bfnm{Rachid}\binits{R.}}
(\byear{2007}).
\btitle{Regularized, fast, and robust analytical Q-ball imaging}.
\bjournal{Magnetic Resonance in Medicine}
\bvolume{58}
\bpages{497--510}.
\end{barticle}
\endbibitem

\bibitem[\protect\citeauthoryear{Fan and Gijbels}{1996}]{Fan-Gijbels96}
\begin{bbook}[author]
\bauthor{\bsnm{Fan},~\bfnm{Jianquing}\binits{J.}} \AND
  \bauthor{\bsnm{Gijbels},~\bfnm{I{\`e}ne}\binits{I.}}
(\byear{1996}).
\btitle{Local Polynomial Modelling and Its Applications}.
\bpublisher{Chapman and Hall}, \baddress{London}.
\end{bbook}
\endbibitem

\bibitem[\protect\citeauthoryear{Fillard {\it
  et~al.}}{2007}]{Fillard-Pennec-Arsigny07}
\begin{barticle}[author]
\bauthor{\bsnm{Fillard},~\bfnm{Pierre}\binits{P.}},
  \bauthor{\bsnm{Pennec},~\bfnm{Xavier}\binits{X.}},
  \bauthor{\bsnm{Arsigny},~\bfnm{Vincent}\binits{V.}} \AND
  \bauthor{\bsnm{Ayache},~\bfnm{Nicholas}\binits{N.}}
(\byear{2007}).
\btitle{Clinical DT-MRI estimation, smoothing, and fiber tracking with
  log-Euclidean metrics}.
\bjournal{Medical Imaging, IEEE Transactions on}
\bvolume{26}
\bpages{1472--1482}.
\end{barticle}
\endbibitem

\bibitem[\protect\citeauthoryear{Fletcher and Joshi}{2007}]{Fletcher-Joshi07}
\begin{barticle}[author]
\bauthor{\bsnm{Fletcher},~\bfnm{P~Thomas}\binits{P.~T.}} \AND
  \bauthor{\bsnm{Joshi},~\bfnm{Sarang}\binits{S.}}
(\byear{2007}).
\btitle{Riemannian geometry for the statistical analysis of diffusion tensor
  data}.
\bjournal{Signal Processing}
\bvolume{87}
\bpages{250--262}.
\end{barticle}
\endbibitem

\bibitem[\protect\citeauthoryear{Friman, Farneback and
  Westin}{2006}]{Friman-Farneback-Westin06}
\begin{barticle}[author]
\bauthor{\bsnm{Friman},~\bfnm{Ola}\binits{O.}},
  \bauthor{\bsnm{Farneback},~\bfnm{Gunnar}\binits{G.}} \AND
  \bauthor{\bsnm{Westin},~\bfnm{Carl-Fredrik}\binits{C.-F.}}
(\byear{2006}).
\btitle{A Bayesian approach for stochastic white matter tractography}.
\bjournal{Medical Imaging, IEEE Transactions on}
\bvolume{25}
\bpages{965--978}.
\end{barticle}
\endbibitem

\bibitem[\protect\citeauthoryear{Gudbjartsson and
  Patz}{1995}]{Gudbjartsson-Patz95}
\begin{barticle}[author]
\bauthor{\bsnm{Gudbjartsson},~\bfnm{H{\'a}kon}\binits{H.}} \AND
  \bauthor{\bsnm{Patz},~\bfnm{Samuel}\binits{S.}}
(\byear{1995}).
\btitle{The Rician distribution of noisy MRI data}.
\bjournal{Magnetic Resonance in Medicine}
\bvolume{34}
\bpages{910--914}.
\end{barticle}
\endbibitem

\bibitem[\protect\citeauthoryear{Hosey, Williams and
  Ansorge}{2005}]{Hosey-Williams-Ansorge05}
\begin{barticle}[author]
\bauthor{\bsnm{Hosey},~\bfnm{Tim}\binits{T.}},
  \bauthor{\bsnm{Williams},~\bfnm{Guy}\binits{G.}} \AND
  \bauthor{\bsnm{Ansorge},~\bfnm{Richard}\binits{R.}}
(\byear{2005}).
\btitle{Inference of multiple fiber orientations in high angular resolution
  diffusion imaging}.
\bjournal{Magnetic Resonance in Medicine}
\bvolume{54}
\bpages{1480--1489}.
\end{barticle}
\endbibitem

\bibitem[\protect\citeauthoryear{Kaufman and
  Rousseeuw}{1990}]{Kaufman-Rousseeuw90}
\begin{bbook}[author]
\bauthor{\bsnm{Kaufman},~\bfnm{Leonard}\binits{L.}} \AND
  \bauthor{\bsnm{Rousseeuw},~\bfnm{Peter~J.}\binits{P.~J.}}
(\byear{1990}).
\btitle{Finding groups in data: an introduction to cluster analysis}
\bvolume{344}.
\bpublisher{John Wiley \& Sons}, \baddress{New Jersey}.
\end{bbook}
\endbibitem

\bibitem[\protect\citeauthoryear{Koch, Norris and
  Hund-Georgiadis}{2002}]{Koch-Norris-Hund-Georgiadis02}
\begin{barticle}[author]
\bauthor{\bsnm{Koch},~\bfnm{Martin~A}\binits{M.~A.}},
  \bauthor{\bsnm{Norris},~\bfnm{David~G}\binits{D.~G.}} \AND
  \bauthor{\bsnm{Hund-Georgiadis},~\bfnm{Margret}\binits{M.}}
(\byear{2002}).
\btitle{An investigation of functional and anatomical connectivity using
  magnetic resonance imaging}.
\bjournal{Neuroimage}
\bvolume{16}
\bpages{241--250}.
\end{barticle}
\endbibitem

\bibitem[\protect\citeauthoryear{Mori}{2007}]{Mori07}
\begin{bbook}[author]
\bauthor{\bsnm{Mori},~\bfnm{Susumu}\binits{S.}}
(\byear{2007}).
\btitle{Introduction to diffusion tensor imaging}.
\bpublisher{Elsevier}, \baddress{Amsterdam}.
\end{bbook}
\endbibitem

\bibitem[\protect\citeauthoryear{Mori and van Zijl}{2002}]{Mori-Zijl02}
\begin{barticle}[author]
\bauthor{\bsnm{Mori},~\bfnm{Susumu}\binits{S.}} \AND \bauthor{\bparticle{van
  }\bsnm{Zijl},~\bfnm{Peter}\binits{P.}}
(\byear{2002}).
\btitle{Fiber tracking: principles and strategies--a technical review}.
\bjournal{NMR in Biomedicine}
\bvolume{15}
\bpages{468--480}.
\end{barticle}
\endbibitem

\bibitem[\protect\citeauthoryear{Mori {\it et~al.}}{1999}]{Mori-Crain-Chacko99}
\begin{barticle}[author]
\bauthor{\bsnm{Mori},~\bfnm{Susumu}\binits{S.}},
  \bauthor{\bsnm{Crain},~\bfnm{Barbara~J}\binits{B.~J.}},
  \bauthor{\bsnm{Chacko},~\bfnm{VP}\binits{V.}} \AND
  \bauthor{\bsnm{Van~Zijl},~\bfnm{Peter}\binits{P.}}
(\byear{1999}).
\btitle{Three-dimensional tracking of axonal projections in the brain by
  magnetic resonance imaging}.
\bjournal{Annals of neurology}
\bvolume{45}
\bpages{265--269}.
\end{barticle}
\endbibitem

\bibitem[\protect\citeauthoryear{Mukherjee {\it
  et~al.}}{2008}]{Mukherjee-Berman-Chung08}
\begin{barticle}[author]
\bauthor{\bsnm{Mukherjee},~\bfnm{P}\binits{P.}},
  \bauthor{\bsnm{Berman},~\bfnm{JI}\binits{J.}},
  \bauthor{\bsnm{Chung},~\bfnm{SW}\binits{S.}},
  \bauthor{\bsnm{Hess},~\bfnm{CP}\binits{C.}} \AND
  \bauthor{\bsnm{Henry},~\bfnm{RG}\binits{R.}}
(\byear{2008}).
\btitle{Diffusion tensor MR imaging and fiber tractography: theoretic
  underpinnings}.
\bjournal{American journal of neuroradiology}
\bvolume{29}
\bpages{632--641}.
\end{barticle}
\endbibitem

\bibitem[\protect\citeauthoryear{Nimsky, Ganslandt and
  Fahlbusch}{2006}]{nimsky2006implementation}
\begin{barticle}[author]
\bauthor{\bsnm{Nimsky},~\bfnm{Christopher}\binits{C.}},
  \bauthor{\bsnm{Ganslandt},~\bfnm{Oliver}\binits{O.}} \AND
  \bauthor{\bsnm{Fahlbusch},~\bfnm{Rudolf}\binits{R.}}
(\byear{2006}).
\btitle{Implementation of fiber tract navigation}.
\bjournal{Neurosurgery}
\bvolume{58}
\bpages{ONS--292}.
\end{barticle}
\endbibitem

\bibitem[\protect\citeauthoryear{Parker and
  Alexander}{2003}]{Parker-Alexander03}
\begin{binproceedings}[author]
\bauthor{\bsnm{Parker},~\bfnm{Geoff J.~M.}\binits{G.~J.~M.}} \AND
  \bauthor{\bsnm{Alexander},~\bfnm{Daniel~C}\binits{D.~C.}}
(\byear{2003}).
\btitle{Probabilistic Monte Carlo based mapping of cerebral connections
  utilising whole-brain crossing fibre information}.
In \bbooktitle{Information Processing in Medical Imaging}
\bpages{684--695}.
\bpublisher{Springer}.
\end{binproceedings}
\endbibitem

\bibitem[\protect\citeauthoryear{Pennec, Fillard and
  Ayache}{2006}]{Pennec-Fillard-Ayache06}
\begin{barticle}[author]
\bauthor{\bsnm{Pennec},~\bfnm{Xavier}\binits{X.}},
  \bauthor{\bsnm{Fillard},~\bfnm{Pierre}\binits{P.}} \AND
  \bauthor{\bsnm{Ayache},~\bfnm{Nicholas}\binits{N.}}
(\byear{2006}).
\btitle{A Riemannian framework for tensor computing}.
\bjournal{International Journal of Computer Vision}
\bvolume{66}
\bpages{41--66}.
\end{barticle}
\endbibitem

\bibitem[\protect\citeauthoryear{Rousseeuw}{1987}]{Rousseeuw87}
\begin{barticle}[author]
\bauthor{\bsnm{Rousseeuw},~\bfnm{Peter~J.}\binits{P.~J.}}
(\byear{1987}).
\btitle{Silhouettes: a graphical aid to the interpretation and validation of
  cluster analysis}.
\bjournal{Journal of computational and applied mathematics}
\bvolume{20}
\bpages{53--65}.
\end{barticle}
\endbibitem

\bibitem[\protect\citeauthoryear{Scherrer and
  Warfield}{2010}]{Scherrer-Warfield10}
\begin{binproceedings}[author]
\bauthor{\bsnm{Scherrer},~\bfnm{Benoit}\binits{B.}} \AND
  \bauthor{\bsnm{Warfield},~\bfnm{Simon~K}\binits{S.~K.}}
(\byear{2010}).
\btitle{Why multiple b-values are required for multi-tensor models. Evaluation
  with a constrained log-{E}uclidean model}.
In \bbooktitle{2010 IEEE International Symposium on Biomedical Imaging: From
  Nano to Macro}
\bpages{1389--1392}.
\end{binproceedings}
\endbibitem

\bibitem[\protect\citeauthoryear{Schwartzman, Dougherty and
  Taylor}{2008}]{Schwartzman-Dougherty-Taylor08}
\begin{barticle}[author]
\bauthor{\bsnm{Schwartzman},~\bfnm{Armin}\binits{A.}},
  \bauthor{\bsnm{Dougherty},~\bfnm{Robert~F}\binits{R.~F.}} \AND
  \bauthor{\bsnm{Taylor},~\bfnm{Jonathan~E}\binits{J.~E.}}
(\byear{2008}).
\btitle{False discovery rate analysis of brain diffusion direction maps}.
\bjournal{The Annals of Applied Statistics}
\bvolume{2}
\bpages{153--175}.
\end{barticle}
\endbibitem

\bibitem[\protect\citeauthoryear{Schwarz}{1978}]{Schwarz78}
\begin{barticle}[author]
\bauthor{\bsnm{Schwarz},~\bfnm{Gideon}\binits{G.}}
(\byear{1978}).
\btitle{Estimating the dimension of a model}.
\bjournal{The Annals of Statistics}
\bvolume{6}
\bpages{461--464}.
\end{barticle}
\endbibitem

\bibitem[\protect\citeauthoryear{Sporns}{2011}]{Sporns11}
\begin{bbook}[author]
\bauthor{\bsnm{Sporns},~\bfnm{Olaf}\binits{O.}}
(\byear{2011}).
\btitle{Networks of the Brain}.
\bpublisher{The MIT Press}.
\end{bbook}
\endbibitem

\bibitem[\protect\citeauthoryear{Tabelow, Voss and
  Polzehl}{2012}]{Tabelow-Voss-Polzehl12}
\begin{barticle}[author]
\bauthor{\bsnm{Tabelow},~\bfnm{K}\binits{K.}},
  \bauthor{\bsnm{Voss},~\bfnm{HU}\binits{H.}} \AND
  \bauthor{\bsnm{Polzehl},~\bfnm{J}\binits{J.}}
(\byear{2012}).
\btitle{Modeling the orientation distribution function by mixtures of angular
  central Gaussian distributions}.
\bjournal{Journal of neuroscience methods}
\bvolume{203}
\bpages{200--211}.
\end{barticle}
\endbibitem

\bibitem[\protect\citeauthoryear{Tournier {\it
  et~al.}}{2004}]{Tournier-Calamante-Gadian04}
\begin{barticle}[author]
\bauthor{\bsnm{Tournier},~\bfnm{J}\binits{J.}},
  \bauthor{\bsnm{Calamante},~\bfnm{Fernando}\binits{F.}},
  \bauthor{\bsnm{Gadian},~\bfnm{David~G}\binits{D.~G.}},
  \bauthor{\bsnm{Connelly},~\bfnm{Alan}\binits{A.}} \betal{et~al.}
(\byear{2004}).
\btitle{Direct estimation of the fiber orientation density function from
  diffusion-weighted MRI data using spherical deconvolution}.
\bjournal{NeuroImage}
\bvolume{23}
\bpages{1176--1185}.
\end{barticle}
\endbibitem

\bibitem[\protect\citeauthoryear{Tournier {\it
  et~al.}}{2007}]{Tournier-Calamante-Connelly07}
\begin{barticle}[author]
\bauthor{\bsnm{Tournier},~\bfnm{J}\binits{J.}},
  \bauthor{\bsnm{Calamante},~\bfnm{Fernando}\binits{F.}},
  \bauthor{\bsnm{Connelly},~\bfnm{Alan}\binits{A.}} \betal{et~al.}
(\byear{2007}).
\btitle{Robust determination of the fibre orientation distribution in diffusion
  MRI: non-negativity constrained super-resolved spherical deconvolution}.
\bjournal{NeuroImage}
\bvolume{35}
\bpages{1459--1472}.
\end{barticle}
\endbibitem

\bibitem[\protect\citeauthoryear{Tuch}{2002}]{Tuch02}
\begin{bphdthesis}[author]
\bauthor{\bsnm{Tuch},~\bfnm{David~S.}\binits{D.~S.}}
(\byear{2002}).
\btitle{Diffusion MRI of complex tissue structure}
\btype{PhD thesis}, \bschool{Massachusetts Institute of Technology}.
\end{bphdthesis}
\endbibitem

\bibitem[\protect\citeauthoryear{Tuch}{2004}]{Tuch04}
\begin{barticle}[author]
\bauthor{\bsnm{Tuch},~\bfnm{David~S.}\binits{D.~S.}}
(\byear{2004}).
\btitle{Q-ball imaging}.
\bjournal{Magnetic Resonance in Medicine}
\bvolume{52}
\bpages{1358--1372}.
\end{barticle}
\endbibitem

\bibitem[\protect\citeauthoryear{Tuch {\it
  et~al.}}{2002}]{Tuch-Reese-Wiegell02}
\begin{barticle}[author]
\bauthor{\bsnm{Tuch},~\bfnm{David~S}\binits{D.~S.}},
  \bauthor{\bsnm{Reese},~\bfnm{Timothy~G}\binits{T.~G.}},
  \bauthor{\bsnm{Wiegell},~\bfnm{Mette~R}\binits{M.~R.}},
  \bauthor{\bsnm{Makris},~\bfnm{Nikos}\binits{N.}},
  \bauthor{\bsnm{Belliveau},~\bfnm{John~W}\binits{J.~W.}} \AND
  \bauthor{\bsnm{Wedeen},~\bfnm{Van~J}\binits{V.~J.}}
(\byear{2002}).
\btitle{High angular resolution diffusion imaging reveals intravoxel white
  matter fiber heterogeneity}.
\bjournal{Magnetic Resonance in Medicine}
\bvolume{48}
\bpages{577--582}.
\end{barticle}
\endbibitem

\bibitem[\protect\citeauthoryear{Weinstein, Kindlmann and
  Lundberg}{1999}]{Weinstein-Kindlmann-Lundberg99}
\begin{binproceedings}[author]
\bauthor{\bsnm{Weinstein},~\bfnm{David}\binits{D.}},
  \bauthor{\bsnm{Kindlmann},~\bfnm{Gordon}\binits{G.}} \AND
  \bauthor{\bsnm{Lundberg},~\bfnm{Eric}\binits{E.}}
(\byear{1999}).
\btitle{Tensorlines: Advection-diffusion based propagation through diffusion
  tensor fields}.
In \bbooktitle{Proceedings of the conference on Visualization}
\bpages{249--253}.
\end{binproceedings}
\endbibitem

\bibitem[\protect\citeauthoryear{Wiegell, Larsson and
  Wedeen}{2000}]{Wiegell-Larsson-Wedeen00}
\begin{barticle}[author]
\bauthor{\bsnm{Wiegell},~\bfnm{Mette~R}\binits{M.~R.}},
  \bauthor{\bsnm{Larsson},~\bfnm{Henrik~BW}\binits{H.~B.}} \AND
  \bauthor{\bsnm{Wedeen},~\bfnm{Van~J}\binits{V.~J.}}
(\byear{2000}).
\btitle{Fiber crossing in human brain depicted with diffusion tensor MR
  imaging1}.
\bjournal{Radiology}
\bvolume{217}
\bpages{897--903}.
\end{barticle}
\endbibitem

\bibitem[\protect\citeauthoryear{Yuan {\it et~al.}}{2012}]{Yuan-Zhu-Lin12}
\begin{barticle}[author]
\bauthor{\bsnm{Yuan},~\bfnm{Ying}\binits{Y.}},
  \bauthor{\bsnm{Zhu},~\bfnm{Hongtu}\binits{H.}},
  \bauthor{\bsnm{Lin},~\bfnm{Weili}\binits{W.}} \AND
  \bauthor{\bsnm{Marron},~\bfnm{J.~S.}\binits{J.~S.}}
(\byear{2012}).
\btitle{Local polynomial regression for symmetric positive definite matrices}.
\bjournal{Journal of the Royal Statistical Society: Series B (Statistical
  Methodology)}
\bvolume{74}
\bpages{697--719}.
\end{barticle}
\endbibitem

\bibitem[\protect\citeauthoryear{Zhu {\it et~al.}}{2007}]{Zhu-Zhang-Ibrahim07}
\begin{barticle}[author]
\bauthor{\bsnm{Zhu},~\bfnm{Hongtu}\binits{H.}},
  \bauthor{\bsnm{Zhang},~\bfnm{Heping}\binits{H.}},
  \bauthor{\bsnm{Ibrahim},~\bfnm{Joseph~G}\binits{J.~G.}} \AND
  \bauthor{\bsnm{Peterson},~\bfnm{Bradley~S}\binits{B.~S.}}
(\byear{2007}).
\btitle{Statistical analysis of diffusion tensors in diffusion-weighted
  magnetic resonance imaging data}.
\bjournal{Journal of the American Statistical Association}
\bvolume{102}
\bpages{1085--1102}.
\end{barticle}
\endbibitem

\end{thebibliography}

\end{document}